\documentclass[obeyspaces,
  journal=pasa,
  manuscript=research-paper, 
  year=2025,
  volume=YY,
]{cup-journal}

\usepackage{amsmath}
\usepackage{amssymb,microtype,siunitx,booktabs}
\usepackage{rotating}
\usepackage{tabularx}

\usepackage[skip=0.5ex]{subcaption}
\usepackage{url}
\usepackage{hyperref} 

\title{Quantifying Radio Source Morphology}

\author{Lachlan J. Barnes}
\affiliation{School of Mathematical and Physical Sciences, 12 Wally's Walk, Macquarie University, NSW 2109, Australia}
\email[Lachlan Barnes]{lachlan.barnes3@hdr.mq.edu.au}

\author{Andrew M. Hopkins}
\affiliation{School of Mathematical and Physical Sciences, 12 Wally's Walk, Macquarie University, NSW 2109, Australia}

\author{Lawrence Rudnick}
\affiliation{Minnesota Institute for Astrophysics, 116 Church St. SE, Minneapolis, MN 55455, USA}

\author{Heinz Andernach}
\affiliation{Depto.\ de Astronom{\'{i}}a, DCNE, Univ.\ de Guanajuato, Callej\'on de Jalisco s/n, Guanajuato, CP 36023, GTO, Mexico}

\author{Michael Cowley}
\affiliation{School of Chemistry \& Physics, Faculty of Science, Queensland University of Technology, Brisbane, QLD 4000, Australia}
\alsoaffiliation{University of Southern Queensland, Centre for Astrophysics, West Street, Toowoomba QLD 4350, Australia}

\author{Nikhel Gupta}
\affiliation{CSIRO Space \& Astronomy, Bentley, WA, Australia}

\author{Ray P. Norris}
\affiliation{ATNF, CSIRO Space \& Astronomy, P.O. Box 76, Epping, NSW 1710, Australia}
\alsoaffiliation{Western Sydney University, Locked Bag 1797, Penrith, NSW 2751, Australia}

\author{Stanislav S. Shabala}
\affiliation{School of Natural Sciences, University of Tasmania, Private Bag 37, Hobart, TAS 7001, Australia}

\author{Tayyaba Zafar}
\affiliation{School of Mathematical and Physical Sciences, 12 Wally's Walk, Macquarie University, NSW 2109, Australia}

\received {10 Mar 2025}
\revised  {06 Jun 2025}
\accepted {26 Jun 2025}
\published{dd Mmm YYYY}

\keywords{radio continuum: galaxies, galaxies: jets, galaxies: structure } 

\begin{document}

\begin{abstract}
The advent of next-generation telescope facilities brings with it an unprecedented amount of data, and the demand for effective tools to process and classify this information has become increasingly important. This work proposes a novel approach to quantify the radio galaxy morphology, through the development of a series of algorithmic metrics that can quantitatively describe the structure of radio source, and can be applied to radio images in an automatic way. These metrics are intuitive in nature and are inspired by the intrinsic structural differences observed between the existing Fanaroff-Riley (FR) morphology types. The metrics are defined in categories of asymmetry, blurriness, concentration, disorder, and elongation ($ABCDE$/single-lobe metrics), as well as the asymmetry and angle between lobes (source metrics). We apply these metrics to a sample of $480$ sources from the Evolutionary Map of the Universe Pilot Survey (EMU-PS) and $72$ well resolved extensively studied sources from An Atlas of DRAGNs, a subset of the revised Third Cambridge Catalogue of Radio Sources (3CRR). We find that these metrics are relatively robust to resolution changes, independent of each other, and measure fundamentally different structural components of radio galaxy lobes. These metrics work particularly well for sources with reasonable signal-to-noise and well separated lobes. We also find that we can recover the original FR classification using probabilistic combinations of our metrics, highlighting the usefulness of our approach for future large data sets from radio sky surveys.
\end{abstract}

\section{Introduction} \label{lit_morph}
\subsection{Radio Galaxies and Morphology}

Supermassive black holes (SMBH) at the core of galaxies, surrounded by an accretion disk of hot gas \citep[e.g.,][]{wilson_difference_1995,best_fundamental_2012,kormendy2013,morganti_radio_2019} are referred to as active galactic nuclei (AGN). AGN sources with radio emission in the form of symmetric jets and lobes are known as DRAGNs (double radio sources associated with a galactic nucleus), a term coined by \citet{leahy_dragns_1993}. DRAGNs (also referred to as ‘radio galaxies’ in this work) exhibit a wide range of structural features and have been classified into distinct morphological types \citep[e.g.,][]{miley_structure_1980}. 

\citet{fanaroff_morphology_1974} first classified DRAGNs into two categories, based on the ratio of $a/b$ \citep[sometimes referred to as the FR ratio, e.g.,][]{brand2023}. Here $a$ represents the distance between brightest spots in opposite lobes, and $b$ is the total extent of the radio emission. Objects with $a/b < 0.5$ were classified as class $1$ (FRI) and those with $a/b > 0.5$ classified as class $2$ (FRII). FRI and FRII type sources appear structurally different. FRI sources tend to have initially collimated jets that decelerate and flare into turbulent, edge-dimmed lobes \citep[e.g.,][]{laingbridle2014,hardcastle2020,saikia_jets_2022}. FRII sources are believed to have highly collimated, relativistic jets that remain narrow over large distances and terminate in bright hotspots at the edges of the lobes, making them edge-brightened \citep[e.g.,][]{shabala_radio_2009,proctor_morphological_2011, saripalli_atlbs_2012,rossi_different_2020,ndungu_classification_2023}. 

The difference between structural features of the two FR types is likely due to both jet power and the radio source environment density \citep[e.g.,][]{best_radio_2009,krause2012,saripalli_atlbs_2012, kapinska_radio_2017, yatesjones2023}. FRI sources tend to be hosted by massive ellipticals in galaxy clusters (i.e., a high density environment), whereas FRII sources tend to be hosted by lower mass galaxies, and are often found in less dense environments \citep{morganti_radio_2019, shabala_duty_2020}. Due to the higher environmental density and lower power, the jets of FRI sources decelerate and dissipate on tens of kpc scales \citep[e.g.,][]{2002MNRAS.336.1161L, laingbridle2014}, whereas those of FRII sources can remain relativistic for hundreds of kpc \citep[e.g.,][]{mingo_revisiting_2019, rossi_different_2020,gordon_quick_2023}. While the original classification indicated that FRII sources tend to have higher luminosities than FRI sources, later work at higher frequencies found there was significant overlap in luminosity for the two morphologies \citep[e.g.,][]{best_radio_2009, gupta_rg-cat_2024}. In addition to luminosity overlap, FRI and FRII sources can also share structural features, making it difficult to precisely determine the morphology of DRAGNs. Low flux density (faint) or sources at different redshifts and environments, can influence the observed morphology leading to hybrid sources \citep[HyMoRS;][]{kapinska_radio_2017}, with one lobe showing FRI-like morphology and the other FRII-like \citep{best_radio_2009,bastien_structured_2021}. Further, \citet{fanaroff_morphology_1974} exclude the central source (core) in their classification. Some authors do not, and this can lead to a core-dominant FRII source being misclassified as an FRI \citep[][]{norris_dragns_25}.

The development of next-generation radio telescope facilities, driven by the vision of the Square Kilometre Array (SKA) since the early 1990s \citep{carilli_motivation_2004}, has led to further advancements in radio telescope technology (such as improved sensitivity, resolution, and survey efficiency). The development of the SKA Pathfinders and their large-scale surveys has resulted in an unprecedented influx of radio data. This has created a growing demand for efficient and effective tools to process and classify radio sources. However, the wide range of morphologies present in radio galaxies makes this a significant challenge for automatic classification. 

\subsection{Machine Learning}

Machine learning (ML) techniques for a wide variety of classification problems have been gradually incorporated into astronomical applications \citep{fayyad_automated_1995}. For example, ML has been used for learning and classifying the morphology of different radio objects \citep[e.g.,][]{ma_machine_2019, wu_radio_2019, galvin_cataloguing_2020, gupta_discovery_2022, gupta_deep_2023, alam2025_SOM}, simulating radio sources \citep[e.g.,][]{bastien_structured_2021, andrianomena_radio_2024} and the development of radio source finding tools \citep[e.g.,][]{riggi_astronomical_2023}. While ML is extremely capable at handling vast amounts of data, the process of training and testing in many ML approaches is computationally intensive \citep[see][]{ball_data_2010}, and the results can be subject to overfitting and generalisation issues \citep{tolley_wavelet_2024}. Further, many of the existing ML models are trained on radio images from a selection of surveys with varying resolution and signal-to-noise which may not be well-matched to the datasets they are later applied to \citep[see][]{ndungu_advances_2023, riggi_astronomical_2023}. The lack of large annotated training datasets remains one of the greatest challenges in assessing and improving the overall performance of ML image classification algorithms \citep{becker_cnn_2021}. Automating the detection and annotation of radio sources is a separate challenge, although recent work \citep{gupta_deep_2023, gupta_radiogalaxynet_2024} introduces a novel ML tool which appears to reliably address this aspect of the problem.

\subsection{Feature Extraction}

Algorithmic feature extraction from images can be faster and more computationally efficient, when compared to many ML approaches, and has been used for image classification across many disciplines for several decades \citep[e.g.,][]{xu_image_2023}. Within astronomy, it has been shown to be a robust approach for classifying the morphology of optical galaxies \citep[e.g.,][]{kent_ccd_1985, abraham_morphologies_1994, bershady_structural_2000}. For example, the development of the CAS parameters \citep[Concentration, Asymmetry, and Smoothness,][]{conselice_asymmetry_2000, conselice_relationship_2003} and the Gini and M20 parameters \citep{lotz_new_2004} have been used extensively to distinguish between early- and late-type galaxies, and have been modified in recent years to be more effective \citep[see][]{ferrari_morfometrykanew_2015}. A number of different tools have also been developed to apply algorithmic feature extraction for the purpose of classifying classes of objects from images. Some examples are WND-CHARM \citep[Weighted Neighbour Distances using a Compound Hierarchy of Algorithms Representing Morphology,][]{orlov_wnd-charm_2008, shamir_automatic_2009} and COSFIRE filters \citep[Combination Of Shifted FIlter REsponses,][]{azzopardi_trainable_2013, ndungu_classification_2023}.

While it is evident that feature extraction methods are powerful tools in the classification and analysis of galaxy morphology at optical wavelengths, their application to radio astronomy has only begun in recent years. Feature extraction techniques have been used to measure the coarse-grained complexity of complex and analogous radio sources \citep[e.g.,][]{segal_identifying_2019, segal_identifying_2023}, and directly analyse radio galaxy features such as FR classification, hotspot brightness \citep[e.g.,][]{becker_classification_2019}, size, eccentricity, orientation, and symmetry of a radio galaxy \citep[e.g.,][]{javaherian_morphological-based_2023}. Other works have applied existing feature based metrics to radio galaxies. For example, \citet{sadeghi_morphological-based_2021} incorporated rotation, translation, and scale invariant image moments, based on Zernike polynomials \citep{teague_image_1980} and \citet{ntwaetsile_rapid_2021} used Haralick features \citep{haralick_textural_1973} to characterise the morphology of FRI and FRII type sources. \citet{brand2023} extracted the FR ratio, if a galaxy is bent or not, the number of bright spots present in the galaxy, and the ratio of the size of bright spots to the total size of the galaxy to aid the training of a CNN in the classification of radio galaxy morphology. Although still incorporating aspects of ML approaches for radio source morphology classification, these studies highlight the potential of algorithmic feature extraction techniques to classify and quantify radio source morphology.

The motivation for this work stems from the need for efficient and scalable methods to classify complex extended radio sources. Algorithmic feature extraction approaches have demonstrated strong potential, and are likely to serve as a valuable method for automating radio galaxy classification. We propose a novel approach to the classification of such sources through the development of a series of algorithmic metrics that can be applied to any radio image. These metrics are designed to quantify key structural features of radio galaxy lobes, such as the concentration of flux or the asymmetry. We seek to establish a foundation for a robust classification framework that can accommodate future radio surveys and higher-resolution observational datasets. Our approach is adaptable to evolving interpretations of radio galaxy morphology. Unlike ML models, which require retraining when new types of sources emerge, our metrics remain agnostic to specific classification schemes and can be applied to any radio galaxy image in an automatic way. 

\section{Data and Pre-processing}

\subsection{EMU and RadioGalaxyNET}

The Evolutionary Map of the Universe \citep[EMU;][]{norris_emu_2011, hopkins25} is a wide-field radio continuum survey, using the ASKAP telescope. The primary goal of EMU is to make a deep ($\sim\! 900$\,MHz) radio continuum survey of the entire Southern sky at a resolution of $\sim\! 15''$ and sensitivity of $\sim\! 25\,\mu$Jy/beam. It is anticipated that EMU will detect and catalogue tens of millions sources, many of which will be extended and complex. A key motivation for this work is the analysis of large numbers of radio sources, for which we start by drawing from EMU data. In preparation for the full EMU survey \citep{norris_emu_2011}, the EMU Pilot Survey \citep[EMU-PS;][]{norris_evolutionary_2021} was conducted to test the survey strategy and data processing pipeline. Phase $1$ of the EMU-PS was limited to a contiguous region of $270~\text{deg}^2$ within the Dark Energy Survey region \citep[DES;][]{abbott_dark_2018}, lying at a declination $<-30^\circ$ and Galactic latitude of $>+20^\circ$, with a sensitivity of $25$–$35\,\mu$Jy/beam at $944$\,MHz.

RadioGalaxyNET \citep{gupta_radiogalaxynet_2024} is developed from a curated dataset of DRAGNs classified by professional astronomers, designed for training machine learning models in radio morphology recognition. Because of the manual nature of identifying the original dataset, the DRAGNs used will tend to favour higher signal-to-noise systems, although this is not a fixed or quantitative threshold. It simply arises since very low surface brightness objects are harder to visually identify. The full RadioGalaxyNET dataset contains $2800$ EMU-PS $15' \times 15'$ cutouts in three channels (unprocessed radio, processed radio, and infrared), resulting in a total of $4155$ annotated sources, since some cutouts contain multiple sources. The annotations consist of radio morphological class information, bounding box details for associated components of each radio galaxy, radio galaxy segmentation masks, and the host galaxy positions from the infrared images. This work primarily analyses a subset of the sources from the RadioGalaxyNET data. 

In order to ensure a source is sufficiently resolved and contains enough observable structure for our analysis, we define a minimum area threshold of the radio cutouts of $20$ times the ASKAP synthesised beam size \citep[$1$ beam $\approx 13''\times11''$;][]{norris_evolutionary_2021}. The minimum area for each cutout is therefore $2860$ arcsec$^{2}$ or $\sim\!715$ pixels ($1$ pixel $\approx 2$ arcsec in the EMU-PS mosaic). Below this value the analysis of the radio lobes can become compromised by the limited number of beams (resolution elements). To produce the cutout images from the EMU-PS, we first extracted the coordinates of the bounding boxes from the RadioGalaxyNET data. These coordinates were transformed to match the orientation of the EMU-PS mosaic, from which we directly produce the cutouts of the individual radio galaxies. There are $595$ sources from RadioGalaxyNET that met the minimum size requirements, with an average cutout size of $\sim\! 3600$ pixels. The cutouts were made on the `unsmooth' EMU-PS mosaic, in order to maximise the resolution in different regions.

\subsection{3CRR and An Atlas of DRAGNs}\label{3c_data}

The Third Cambridge Catalogue of Radio Sources \citep[3C;][]{edge_survey_1959, archer_studies_1959} is a radio catalogue originally observed at $159$\,MHz, then revised at $178$\,MHz \citep[3CR;][]{bennett_preparation_1962, bennett_revised_1962}. The 3CR catalogue contained $328$ sources and was one of the first comprehensive lists of extragalactic radio sources. However, the original survey had relatively limited angular resolution, leading to source blending and other biases \citep{veron_3c}. To improve upon this, a further revision of $173$ bright radio sources at $178$\,MHz was conducted \citep[3CRR;][]{laing_bright_1983}. While the initial 3CR selection suffered from resolution limitations, 3CRR sources have since been observed in much greater detail and resolution. The 3CRR catalogue contains all extragalactic radio sources in the Northern Hemisphere with a $178$\,MHz flux density greater than $10.9$\,Jy. 

\begin{figure}[h]
    \centering
    \includegraphics[width=\linewidth]{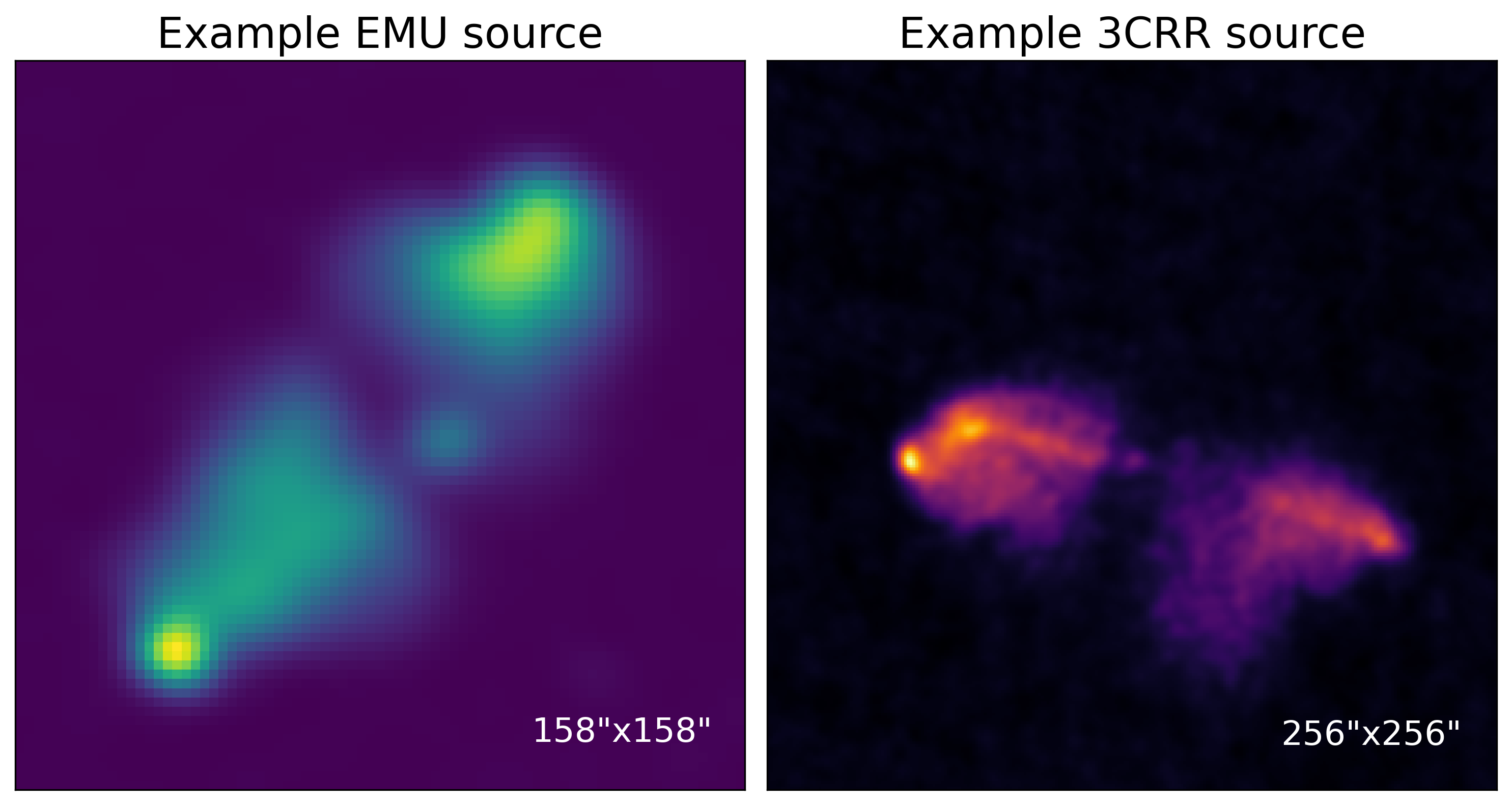}
    \caption{Comparison of two double lobed radio sources to highlight the resolution between datasets. As observed in J202644.4$-$552227 (left) and 3C285 (right), both sources show similar structural features. Angular size of each image shown in the bottom right.}
    \label{3c_emu_comp}
\end{figure}

An Atlas of DRAGNs \citep{leahy_atlas_2013} is a subset of the 3CRR catalogue, consisting of the nearest $85$ DRAGNs. The Atlas of DRAGNs website\footnote{\hyperlink{https://www.jb.man.ac.uk/atlas/}{https://www.jb.man.ac.uk/atlas/}} contains high-quality images of the DRAGNs compiled from different studies. The images in the atlas have high resolution, and sufficient sensitivity and spatial frequency coverage to clearly and accurately image the faintest regions of a radio source. Most images are at frequencies near $1.5$\,GHz, although the frequencies range from $0.3$ to $8.4$\,GHz in the atlas. The difference in resolution between EMU-PS and 3CRR sources from this atlas is highlighted in Figure~\ref{3c_emu_comp}. There is also a dedicated page providing supplementary information about each radio source, containing FR classification, redshift, radio power, and linear size. Given the well-documented nature of these sources, we use this atlas as a reference dataset in this work.

The cutouts of the 3CRR sources, as well as the corresponding supplementary information, were directly downloaded from the Atlas of DRAGNs site. We do not use the full sample of $85$ sources in this work. Through manual inspection, it was found that two sources had corrupt FITS files, two sources were not centred in the image, two were extremely bent (see \ref{trouble_imgs}), and seven contained excessive structural complexity too intricate to be processed reliably in an automatic way (see Table~\ref{tab:omitted_sources}). Our reference dataset therefore consists of $72$ 3CRR sources ($17$ FRI and $55$ FRII). 

\begin{table}[h]
    \centering
    \begin{tabularx}{\linewidth}{p{3cm} X}
        \hline
        \hline
        \textbf{Reason for Omission} & \textbf{Sources} \\
        \hline
        Corrupt FITS & 3C236, NGC6251 \\
        Not centred & 3C35, 3C465 \\
        Extremely bent & 3C83.1B, 3C264 \\
        Complicated structure & 3C16, 3C48, 3C84, 3C274, 3C293, 3C314.1, 3C274 \\
        \hline
        \hline
    \end{tabularx}
    \caption{List of sources and reasons for omission from analysis.}
    \label{tab:omitted_sources}
\end{table}

\section{Development of Quantitative Metrics}\label{development}

As discussed above in Section \ref{lit_morph}, the morphology of a radio source does not always fit into well defined, rigid categories (such as FRI or FRII for example). Further, an object's classification can become ambiguous if it contains characteristics of different categories (e.g., HyMoRS) or when its observed structure varies with resolution. Recently, \citet{rudnick_radio_2021} proposed the concept of assigning radio sources descriptive criteria-based \textit{\#tags} instead of putting sources into such categories. Such \textit{\#tags} must have well defined criteria, and are ideally quantitative. 

Of particular relevance to this work are structural \textit{\#tags}, which contain information about the number of jets, the peak flux separation, FR-morphology, shape, or symmetry, among other properties for a given radio source. In this work, we define a series of metrics that can be used to characterise and quantify the structure of a radio source and be incorporated into the \textit{\#tag} system. These metrics include asymmetry, blurriness, concentration, disorder, and elongation (referred to as the $ABCDE$ metrics), which are applied to a single lobe of a DRAGN system. We also present metrics that are applied to the whole DRAGN system, the asymmetry and angle between lobes (ABL). A brief summary of each of the metrics is provided in Table~\ref{metric_summary}. As the shape of a radio lobe is directly related to the astrophysics of the host AGN, we selected metrics designed to quantitatively capture both morphological structure and flux distribution of a lobe. It is important to note that while many of the following metrics are not new, the innovation in this work lies in their application to quantifying the structure of extended, complex radio sources.

\begin{table}[h]
    \centering
    \begin{tabularx}{\linewidth}{>{\centering\arraybackslash}p{1.8cm} X}
        \hline
        \hline
        \textbf{Metric} & \textbf{Summary} \\
        \hline
        Asymmetry & Multiple metrics. Shape asymmetry, flux distribution asymmetry, shape and flux distribution asymmetry between both lobes of a given source. \\
        Blurriness & Measure of how blurred the edge of a radio lobe is. \\
        Concentration & Concentration of flux within a lobe. \\
        Disorder & Measure of how complicated the shape of the lobe is. \\
        Elongation & Measure of how stretched or compact a radio lobe is. \\
        Angle Between Lobes & Measure of how bent a radio source is. \\
        \hline
        \hline
    \end{tabularx}
    \caption{Summary of the metrics presented in this work. All metrics are unitless except for the angle between lobes, which is measured in degrees.}
    \label{metric_summary}
\end{table}

To calculate each of the above metrics, all radio galaxy source images (from both the EMU-PS and 3CRR catalogue) are divided into two halves, with the expectation that each half will contain one lobe of the galaxy. For rectangular images, we cut the image perpendicular to and half-way along the major axis. For square images, we simply cut the x-axis in half. This is assuming the source is both centred and symmetrically distributed in the image. From a visual inspection of both EMU and 3CRR datasets, the vast majority of galaxies were centred in the radio image. Through manual inspection, we found 3 sources that were extremely bent with both lobes in one half of the cutout (See \ref{trouble_imgs}). These sources were dropped from our analysis. We then identify the brightest pixel in each lobe, ensuring a minimum distance of three pixels from the border of each half. Further, for all metrics except the angle between lobes and asymmetry, we use a minimum threshold of $10\%$ of the peak flux value to ensure lobe isolation and minimise biases from high noise pixels, setting pixel values below this threshold to NaN (Not a Number). For metrics that involve the summation of pixels, the NaNs are excluded. 

\subsection{Pre-processing}\label{preprocessing}

Many of the EMU-PS sources contain a very bright core. As the metric calculations are based on the brightest pixel, this can influence the results. We therefore implement a core-removal process. We first apply a minimum flux threshold of $150\,\mu$Jy, corresponding to roughly $5\sigma$ of the EMU-PS imaging \citep[$\sigma = 25$-$30\,\mu$Jy/beam,][]{norris_evolutionary_2021}. Pixels with intensities less than this threshold are are set to zero to remove noise for the purposes of core identification only. Following the work outlined in \citet{hancock_compact_2012}, to highlight potential edges and features in an image we generate a curvature map by convolving the thresholded image with the following Laplacian kernel:

\begin{equation}
    L^2_{xy} = \begin{bmatrix}
    1 & 1 & 1 \\
    1 & -8 & 1 \\
    1 & 1 & 1 \\
\end{bmatrix}~.
\label{Laplac_kernel}
\end{equation}

\hfill

\begin{figure}[h]
    \centering
    \includegraphics[width=\linewidth]{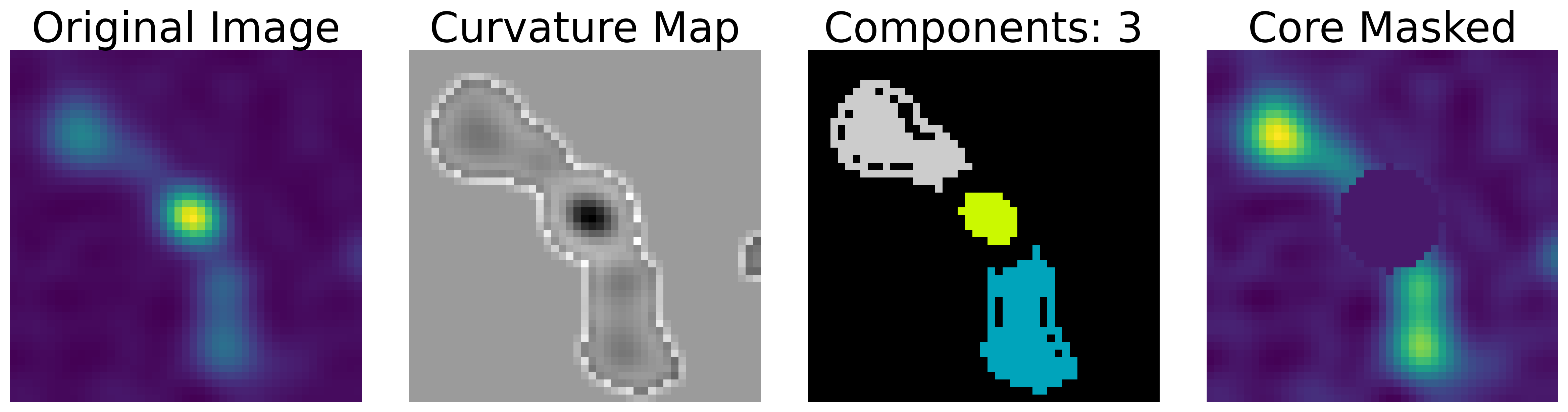}
    \caption{Visualisation of the core masking process. From left to right is the original EMU-PS radio image, the curvature map, the labelled components, and the final image with the core removed. }
    \label{core_mask_fig}
\end{figure}

The resulting curvature map emphasises regions with sharp intensity transitions, aiding in the accurate localisation of a core. We then generate a binary map with labelled features, calculate their sizes, and retain the largest components. If an image contains two lobes and a core (i.e., number of components $> 2$), a circular mask is applied around the local maximum nearest to the infrared coordinates of the host galaxy, which is expected to be coincident with the core. This process is fully automatic and demonstrated in Figure~\ref{core_mask_fig}. Through an initial visual inspection of the $592$ EMU sources, we estimate that $186$ ($30\%$) sources contain a dominant core, i.e., the central component was brighter than the two lobes. We then applied the above core-removal process to the $592$ images and, after another visual inspection, the core of $102$ sources was successfully removed. However, this process is not perfect (see \ref{trouble_imgs}). We found that the number of components detected was sometimes incorrect, leading sources that did not have a dominant core to accidentally have a lobe masked. Sources with a dominant core remaining after the attempted masking ($84$) or an inadvertently masked lobe ($28$) were subsequently dropped from our analysis, leading to a final EMU-PS dataset consisting of $480$ sources.

\subsection{Asymmetry}

The asymmetry of a galaxy has been defined in a number of ways in optical astronomy \citep[e.g.,][]{schade_canada-france_1995, abraham_galaxy_1996, abraham_morphologies_1996, conselice_symmetry_1997, brinchmann_hubble_1998}. These methods provide a framework for morphological classification that can also be applied to radio sources. We adopt the approach to asymmetry as described in \citet{conselice_asymmetry_2000} and \citet{conselice_relationship_2003}, which compares a source image ($I$), with itself rotated by an angle of $180^\circ$ ($I_{Rot}$). We define the centre of rotation as the brightest pixel in a given lobe. We produce the difference image between the original and rotated image, and take the absolute value of the sum over all pixels, $\eta$, and normalising by the total flux of the original lobe, $I_{Tot}$, to produce a shape asymmetry metric ($A_{S}$). The absolute value of the difference of two images will produce twice the flux for a perfectly asymmetric source (i.e., each bright pixel has no corresponding bright pixel after the image is rotated), and we therefore include the factor of $1/2$ to ensure $A_{S}$ is always between $0$ and $1$:

\begin{equation}
    A_{S} = \frac{1}{2} \sum_{i,j}^{\eta} \frac{|I - I_{Rot}|}{I_{Tot}}~.
    \label{shape_res}
\end{equation}

We also include a metric to describe the flux distribution across a lobe, defining a ratio asymmetry ($A_{R}$), which is the absolute value of the ratio of the original and rotated lobe, summed over all pixels, $\eta$ (Equation~\ref{rat_res}). For a completely uniform (or symmetric) source, each pixel ratio would be $1$, and the sum of the ratio image will be number of pixels. We therefore normalise by the number of pixels, meaning a perfectly uniform source will produce a minimum $A_{R}$ value of $1$: 

\begin{equation}
    A_{R} = \frac{1}{\eta} \sum_{i,j}^{\eta} \big| \frac{I}{I_{Rot}} \big|~.
    \label{rat_res}
\end{equation}

\begin{figure}[h]
    \centering
    \includegraphics[width=\linewidth]{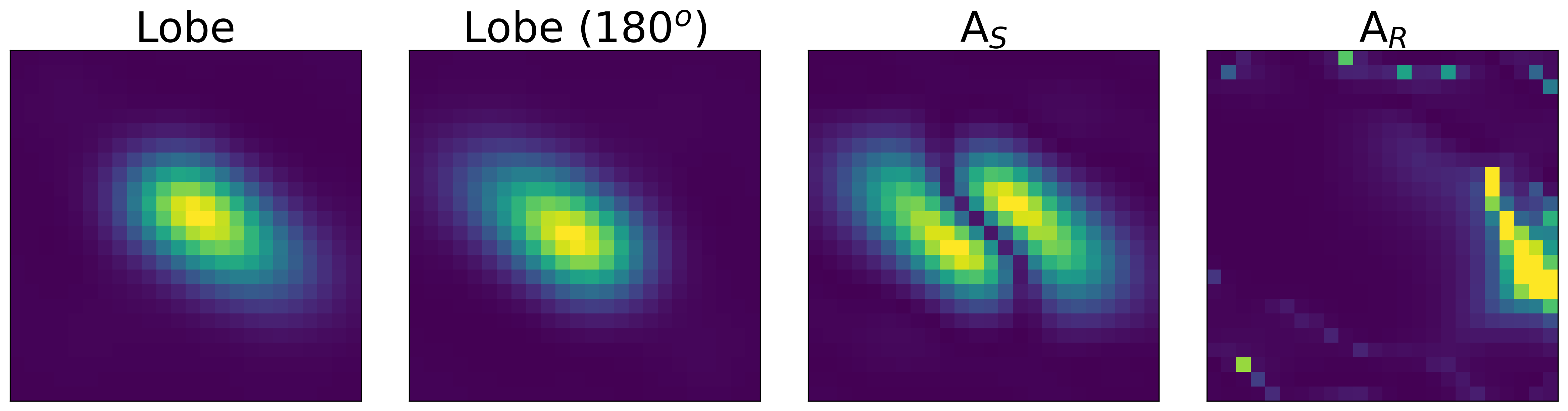}
    \caption{Visualisation of both asymmetry calculations. From left to right: original lobe image, its rotated counterpart, the residual image after taking the absolute value of the difference between the original and rotated image, and the residual image after taking the ratio of the original and rotated images.}
    \label{asymm_eg}
\end{figure}

\noindent Examples of the images of $A_{S}$ and $A_{R}$ (before summing the pixels) are shown in Figure~\ref{asymm_eg}.

We can define analogous calculations to $A_{S}$ and $A_{R}$ to compare how asymmetric the two lobes ($I_{1}$ and $I_{2}$) of a radio galaxy are to each other. We denote these new parameters with the subscript $p$ to represent the comparison of the pair of lobes in the DRAGN system. We can define two shape asymmetry metrics, the first being the total asymmetry ($A_{S,p}$, Equation~\ref{total_res_pair}) which is weighted by the relative flux in each lobe. This takes a similar form to the $A_S$ calculation, however we instead normalise by the sum of the total flux in both lobes. To compare only how symmetric shapes of the lobes are, without the relative flux weighting, we must normalise each lobe by the total flux, $I_{Tot}$ before computing the difference ($A'_{S,p}$, Equation~\ref{shape_res_pair}). The ratio asymmetry between lobes, $A_{R,p}$ (Equation~\ref{tot_rat_res_pair}), is determined in the same way as $A_{R}$, using the absolute value of the ratio of one lobe with the rotated other, normalised by the number of pixels, $\eta$. As before, we can also normalise by the total flux before computing the ratio ($A'_{R,p}$, Equation~\ref{rat_res_pair}). As with the single-lobe calculations, $A_{S,p}$ and $A'_{S,p}$ both have values between $0$ and $1$, and $A_{R,p}$ and $A'_{R,p}$ are always greater than $1$.

\begin{equation}
    A_{S,p} = \frac{1}{2} \sum_{i,j}^{\eta} \frac{|I_{1} - I_{2,Rot}|}{I_{1, Tot} + I_{2, Tot}}~,
    \label{total_res_pair}
\end{equation}

\begin{equation}
    A'_{S,p} = \frac{1}{2} \sum_{i,j}^{\eta} \big|\frac{I_{1}}{I_{1,Tot}} - \frac{I_{2}}{I_{2,Tot}}\big|~,
    \label{shape_res_pair}
\end{equation}

\begin{equation}
    A_{R,p} = \frac{1}{\eta}\sum_{i,j}^{\eta} \big| \frac{I_{1}}{I_{2,rot}} \big|~,
    \label{tot_rat_res_pair}
\end{equation}

\begin{equation}
    A'_{R,p} = \frac{1}{\eta}\sum_{i,j}^{\eta} \big| \frac{I_{1}/I_{1,Tot}}{I_{2,rot}/I_{2,Tot}} \big|~.
    \label{rat_res_pair}
\end{equation}

\subsection{Blurriness}

\begin{figure}[h]
    \centering
    \includegraphics[width=\linewidth]{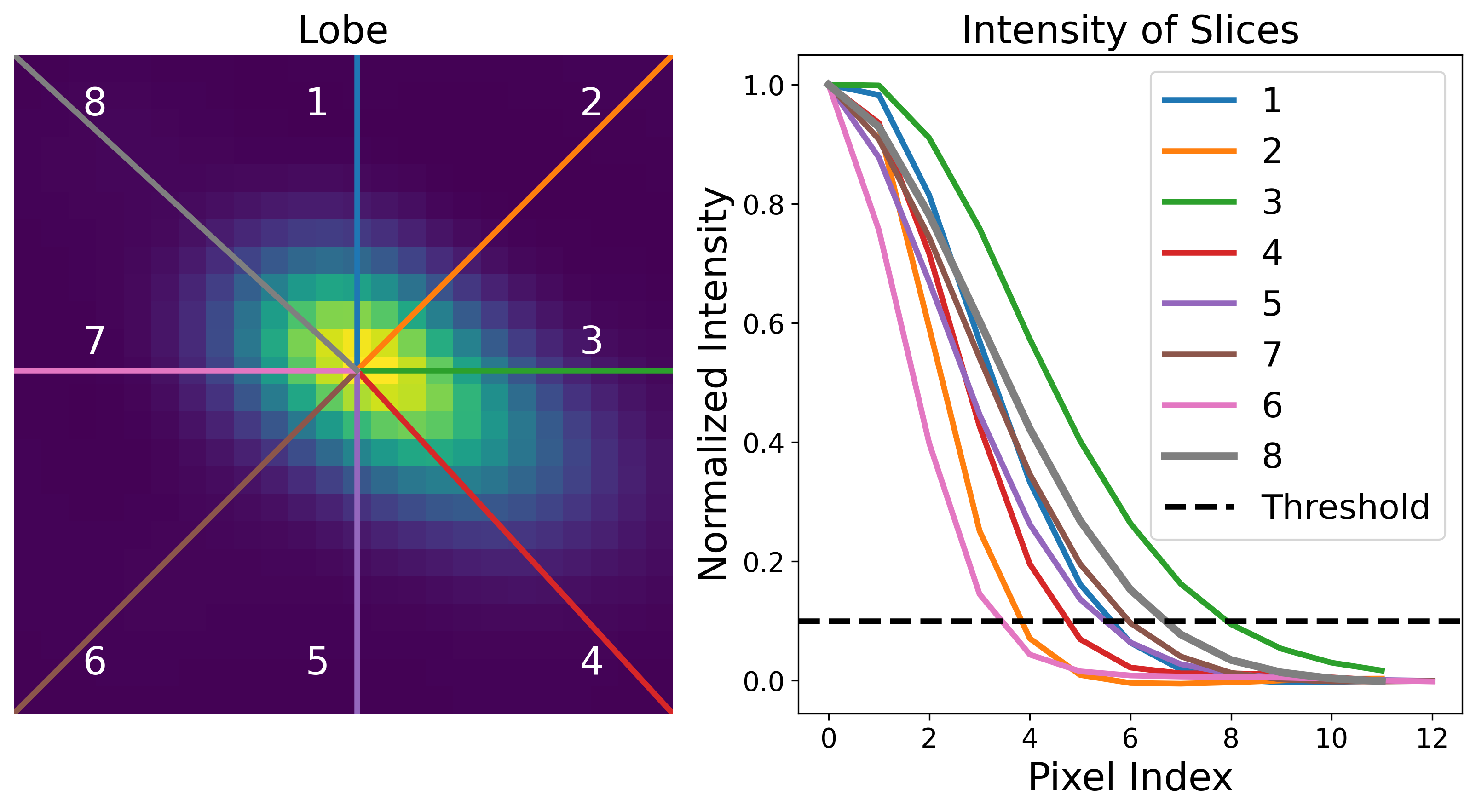}
    \caption{Visualisation of the blurriness calculation, highlighting the 8 cardinal directions and corresponding normalised intensity slices measured away from the brightest pixel of a lobe.}
    \label{gradient_fig}
\end{figure}

We calculate how blurry (or diffuse) radio lobes are by measuring the rate at which the flux of a lobe decreases from the hotspot radially outwards. We first define eight cardinal directions, taking slices of the pixel intensities radially outwards from the brightest pixel (i.e., the hotspot). To calculate the rate at which the flux decreases, we use the \texttt{scipy.optimize curve\_fit} function to fit a straight line to each of the eight slices and extract the gradient (Figure~\ref{gradient_fig}). While we acknowledge that the pixel intensity is not linear, this approach provides a robust first-order estimate of blurriness, as more complex models (e.g., exponential or power law fits) introduce additional parameters that may not be necessary for comparative classification. The gradient of pixel intensities (normalised by the peak brightness) in the whole slice alone is not enough to determine how blurred or sharp a lobe is. We therefore also measure the gradient of the three outermost pixels in each slice. We chose to use 3 pixels as this is the minimum number of points we can have to fit a gradient, and limit the number of pixels from within the lobe. We average over all eight directions for both the full slice and outermost pixels ($\Bar{g}_{\text{full}}$ and $\Bar{g}_{\text{outer}}$, respectively). To estimate the how blurred a lobe is, we then take the ratio of $\Bar{g}_{\text{full}}$ and $\Bar{g}_{\text{outer}}$:

\begin{equation}
    B = \frac{\Bar{g}_{\text{full}}}{\Bar{g}_{\text{outer}}}~.
    \label{gradient_eqn}
\end{equation}

\noindent Small $B$ values (low blurriness) are produced when the average edge gradient of flux is steeper than the average gradient of the whole lobe. This could be indicative of a relatively constant brightness distribution across the lobe with a sharp drop to the background, possibly produced by a lobe expanding into a dense environment, such as within a galaxy cluster. Conversely, higher $B$ values (high blurriness) may indicate more diffuse lobe structure and potentially a less dense environment.

\subsection{Concentration}

The concentration metric describes how the flux is distributed within a lobe. Our approach to concentration is analogous to the approach described in \citet{abraham_morphologies_1994}, with concentration defined as the ratio of flux between inner and outer isophotes of normalised radii. We define concentration, $C$, as the ratio of radii containing $80\%$ and $20\%$ of the flux in a lobe, centred on the brightest pixel:

\begin{equation}
            C = \frac{r_{80}}{r_{20}}~.
            \label{conc_eq}
\end{equation}

\begin{figure}[h]
    \centering
    \includegraphics[width=\linewidth]{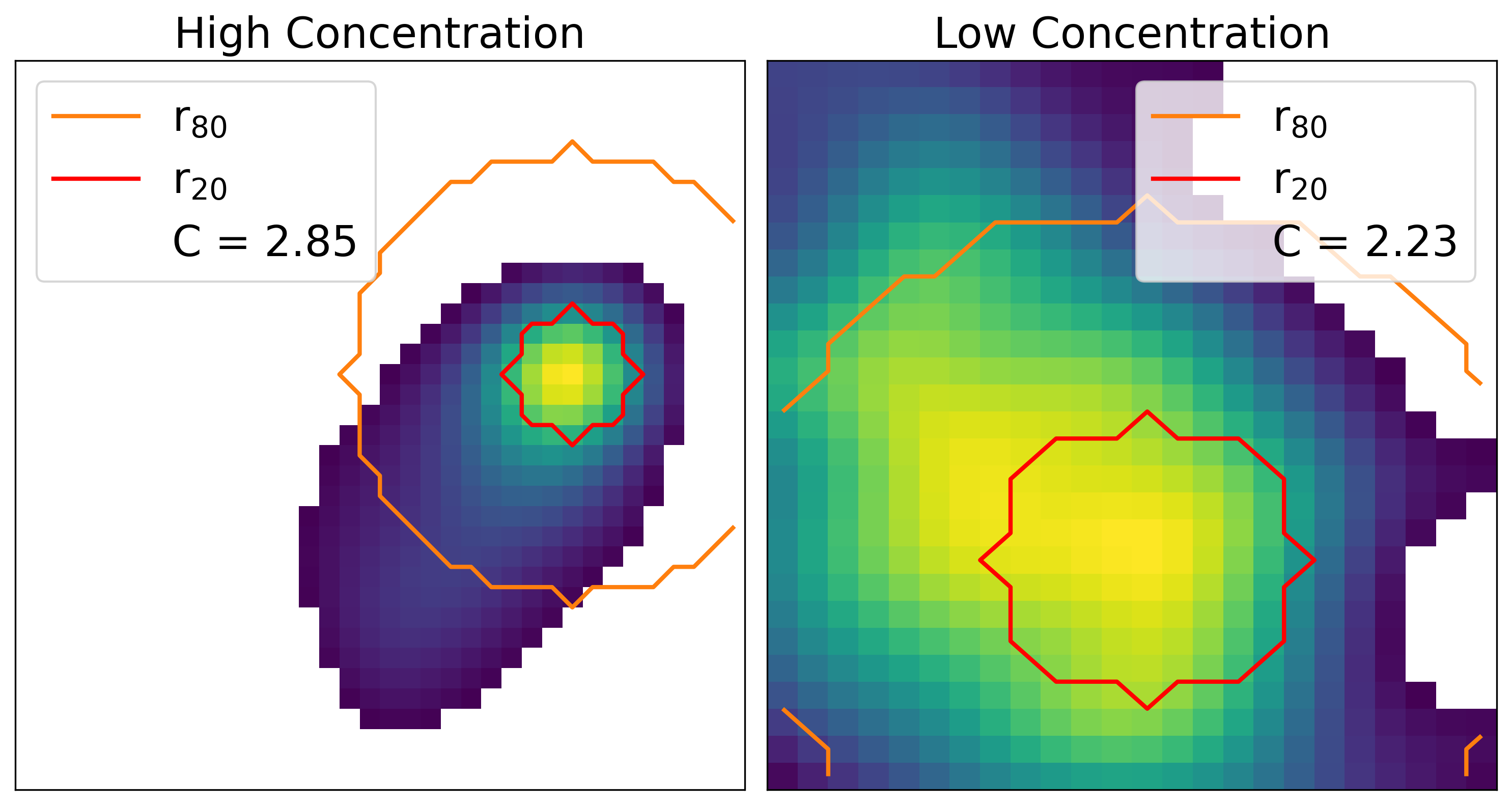}
    \caption{Visualisation of the concentration metric comparing lobes with high and low concentration (left and right, respectively). Radii corresponding to 20\% and 80\% of the lobe flux shown in red and orange, respectively.}
    \label{concentration_fig}
\end{figure}

 To calculate the radius of each isophote, we sum the number of pixels until $20\%$ and $80\%$ of the total lobe flux is achieved. This sum is treated as an area, $\alpha$, and converted into a circular radius, $r$, using the relationship $r = \sqrt{\alpha/\pi}$. Here, $C$ has a minimum value of $2$ for a completely uniform source. Larger values of $C$ mean the flux is more concentrated within the lobe, and can be indicative of a hotspot. A comparison showing high and low concentration lobes is presented in Figure~\ref{concentration_fig}.

\subsection{Disorder}

We include a metric to quantify how perturbed a radio galaxy lobe is. Perturbations in the lobe structure can be caused by differing jet power, or from dynamic forces within the host galaxy environment. We refer to this metric as `disorder', $D$, and following the definition of complexity by \citet{watson_perimetric_2012}, define it as the ratio of the perimeter of a lobe squared and its area multiplied by $4\pi$: 

\begin{equation}
    \text{D} = \frac{\text{Perimeter}^2}{4\pi\text{Area}}~. 
    \label{complex}
\end{equation}

\begin{figure}[h]
    \centering
    \includegraphics[width=\linewidth]{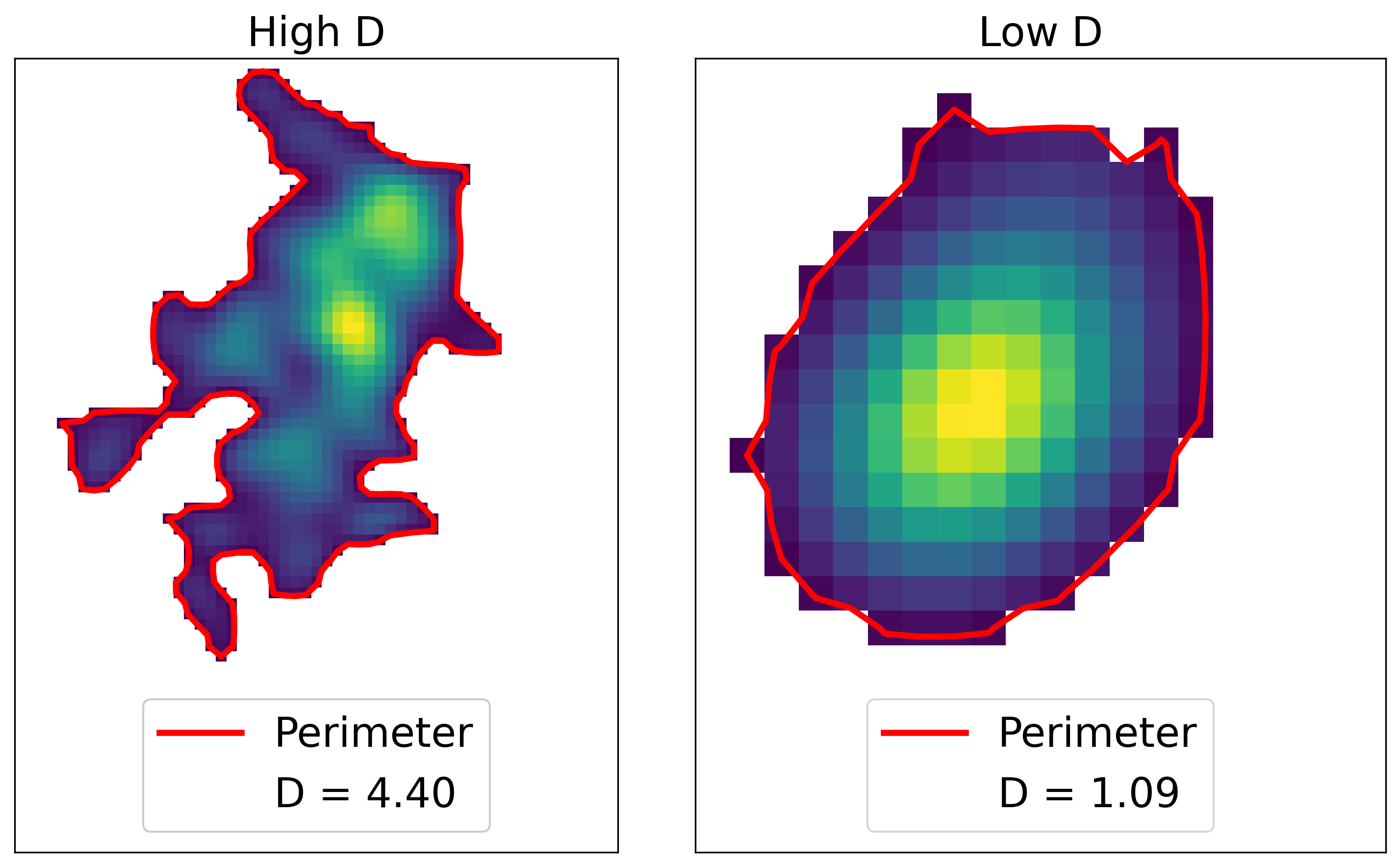}
    \caption{Visualisation of the disorder metric comparing lobes with high and low disorder (left and right, respectively). Perimeter calculated by the length of the red contour.}
    \label{complexity_fig}
\end{figure}

To determine the perimeter, we measure the length of a fitted contour around pixels above the minimum threshold ($10\%$ of the peak brightness). To calculate the area of the lobe, we simply sum the number of non-NaN pixels above the minimum threshold. The $D$ metric therefore has a minimum value of $1$ for a perfectly circular source. Figure~\ref{complexity_fig} highlights the difference between lobes with high and low disorder.

\subsection{Elongation}

We also measure the elongation of a lobe for a radio source, which can be related to the power of a radio jet as more powerful radio jets can produce more extended structures. We define it as the ratio of the major axis and the minor axis of a bounding box fitted to a radio lobe:

\begin{equation}
    \text{E} = \frac{\text{Major axis}}{\text{Minor axis}}~.
    \label{elong}
\end{equation}

\begin{figure}[h]
    \centering
    \includegraphics[width=\linewidth]{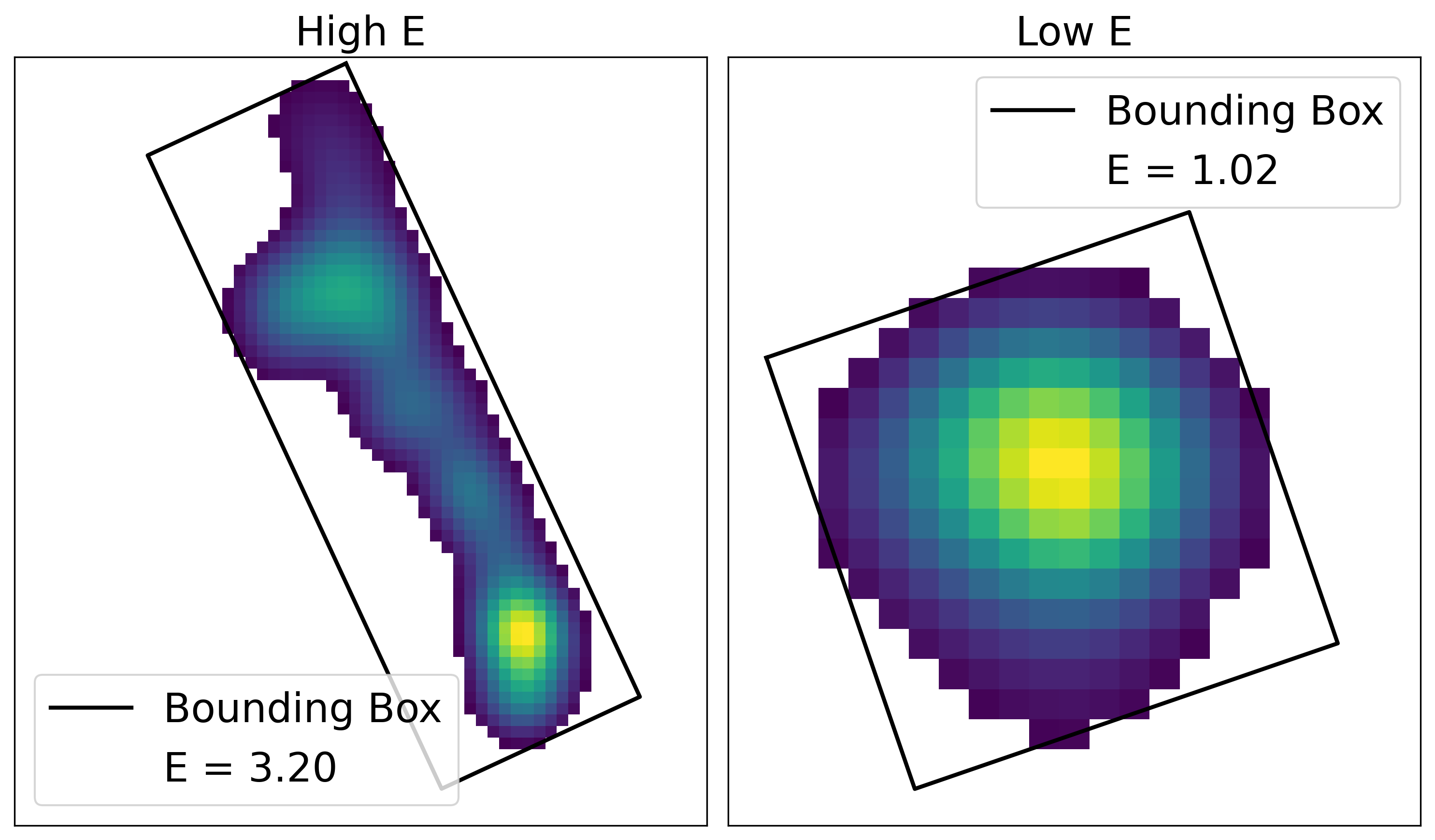}
    \caption{Visualisation of the elongation metric comparing lobes with high and low elongation (left and right, respectively). The bounding box of each lobe is shown in black.}
    \label{elongation_fig}
\end{figure}

This metric will have a minimum value of $1$ for a square or circularly shaped object. We perform a principle component analysis (PCA) on the pixels above the minimum threshold to orient and fit the bounding box, where we can then extract the lengths of the major and minor axes of the box. This is shown in Figure~\ref{elongation_fig}. 

\subsection{Angle Between Lobes}

\begin{figure}[h]
    \centering
    \includegraphics[width=\linewidth]{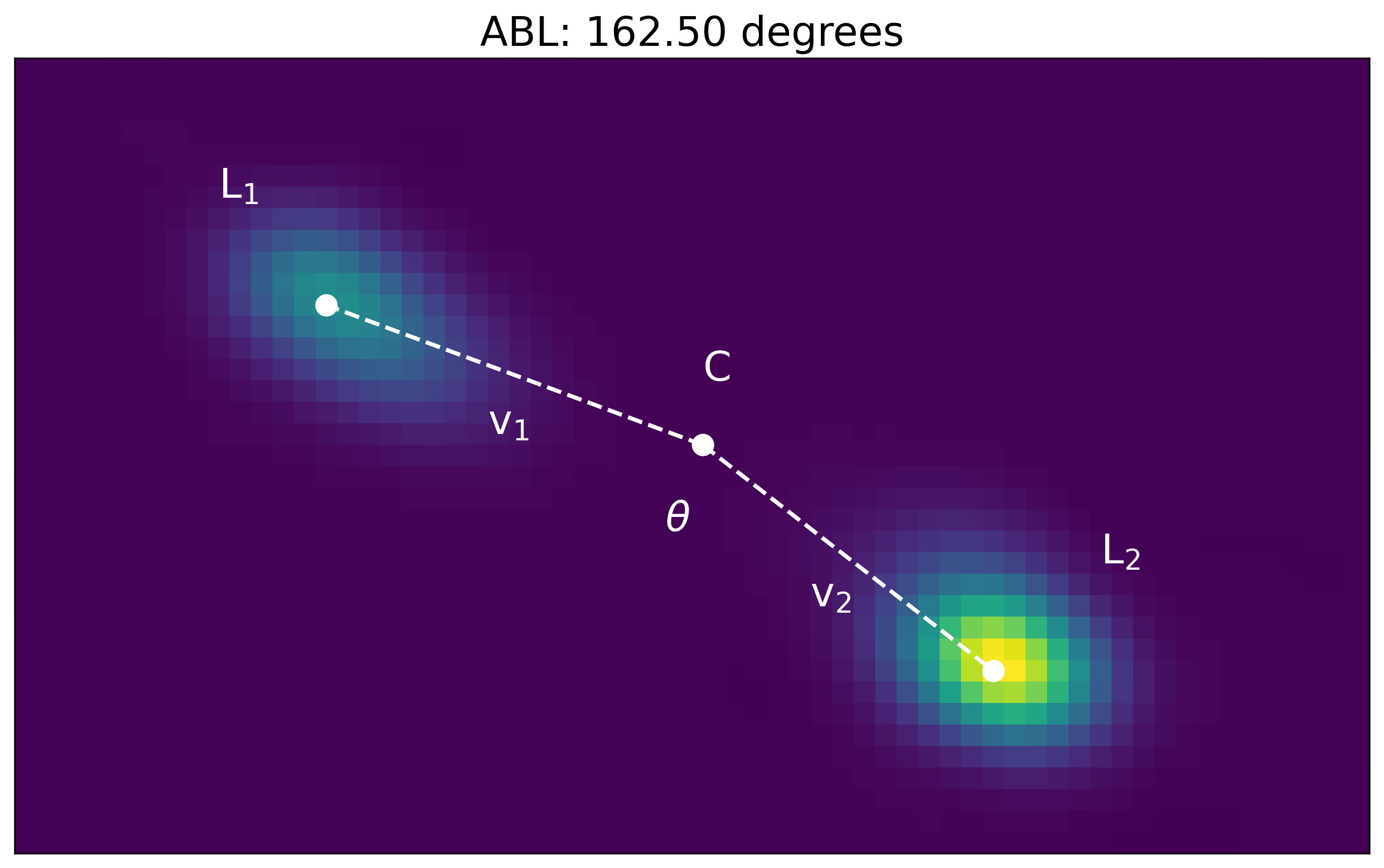}
    \caption{Example calculation of the ABL for a radio source from the EMU-PS. Positions of the brightest pixels in each lobe and the location of the host galaxy shown by the white dots and labelled $\mathbf{L}_1$, $\mathbf{L}_2$ and $\mathbf{C}$, respectively. Vectors $\mathbf{v}_1$ and $\mathbf{v}_2$ shown by the dashed lines. Angle $\theta$ indicated by the arc between both vectors.}
    \label{ABL_fig}
\end{figure}

We include a metric that estimates how bent a radio source is, the angle between lobes (ABL). The ABL can serve as a proxy for environmental effects, as bent radio galaxies are predominantly found in dense and dynamic environments such as galaxy clusters. To calculate the ABL, we treat the brightest pixels in each lobe as points in space, $\mathbf{L}_1$ and $\mathbf{L}_2$, where we can then determine vectors to the core of the radio galaxy, $\mathbf{C}$  (Equations~\ref{eq:v1} and \ref{eq:v2}). For the location of the core, we prioritise using the infrared coordinates of the host galaxy when available, otherwise we simply use the geometric centre of original radio galaxy cutout. Once the vectors have been determined, we compute the dot product (Equation~\ref{eq:dot_product}) and magnitude of the vectors (Equations~\ref{eq:magnitude_v1} and \ref{eq:magnitude_v2}) to extract the angle between lobes (Equation~\ref{eq:theta}), which is then converted into degrees. This calculation is illustrated in Figure~\ref{ABL_fig}. We will refer to ABL, $A_{S,p}$, $A'_{S,p}$, $A_{R,p}$, and $A'_{R,p}$ as source metrics hereafter.

\begin{align}
    \mathbf{v}_1 &= \mathbf{L}_1 - \mathbf{C} = (x_1 - x_c, y_1 - y_c)~, \tag{\theequation a} \label{eq:v1} \\
    \mathbf{v}_2 &= \mathbf{L}_2 - \mathbf{C} = (x_2 - x_c, y_2 - y_c)~, \tag{\theequation b} \label{eq:v2} \\
    \mathbf{v}_1 \cdot \mathbf{v}_2 &= (x_1 - x_c)(x_2 - x_c) + (y_1 - y_c)(y_2 - y_c)~, \label{eq:dot_product} \\
    |\mathbf{v}_1| &= \sqrt{(x_1 - x_c)^2 + (y_1 - y_c)^2}~, \tag{\theequation a} \label{eq:magnitude_v1} \\
    |\mathbf{v}_2| &= \sqrt{(x_2 - x_c)^2 + (y_2 - y_c)^2}~, \tag{\theequation b} \label{eq:magnitude_v2} \\
    \cos(\theta) &= \frac{\mathbf{v}_1 \cdot \mathbf{v}_2}{|\mathbf{v}_1| |\mathbf{v}_2|} \Rightarrow \theta = \arccos\left( \frac{\mathbf{v}_1 \cdot \mathbf{v}_2}{|\mathbf{v}_1| |\mathbf{v}_2|} \right)~.
    \label{eq:theta}
\end{align}

\subsection{Resolution Dependence of Metrics}\label{res_dependence}

\begin{figure}[h]
    \centering
    \includegraphics[width=\linewidth]{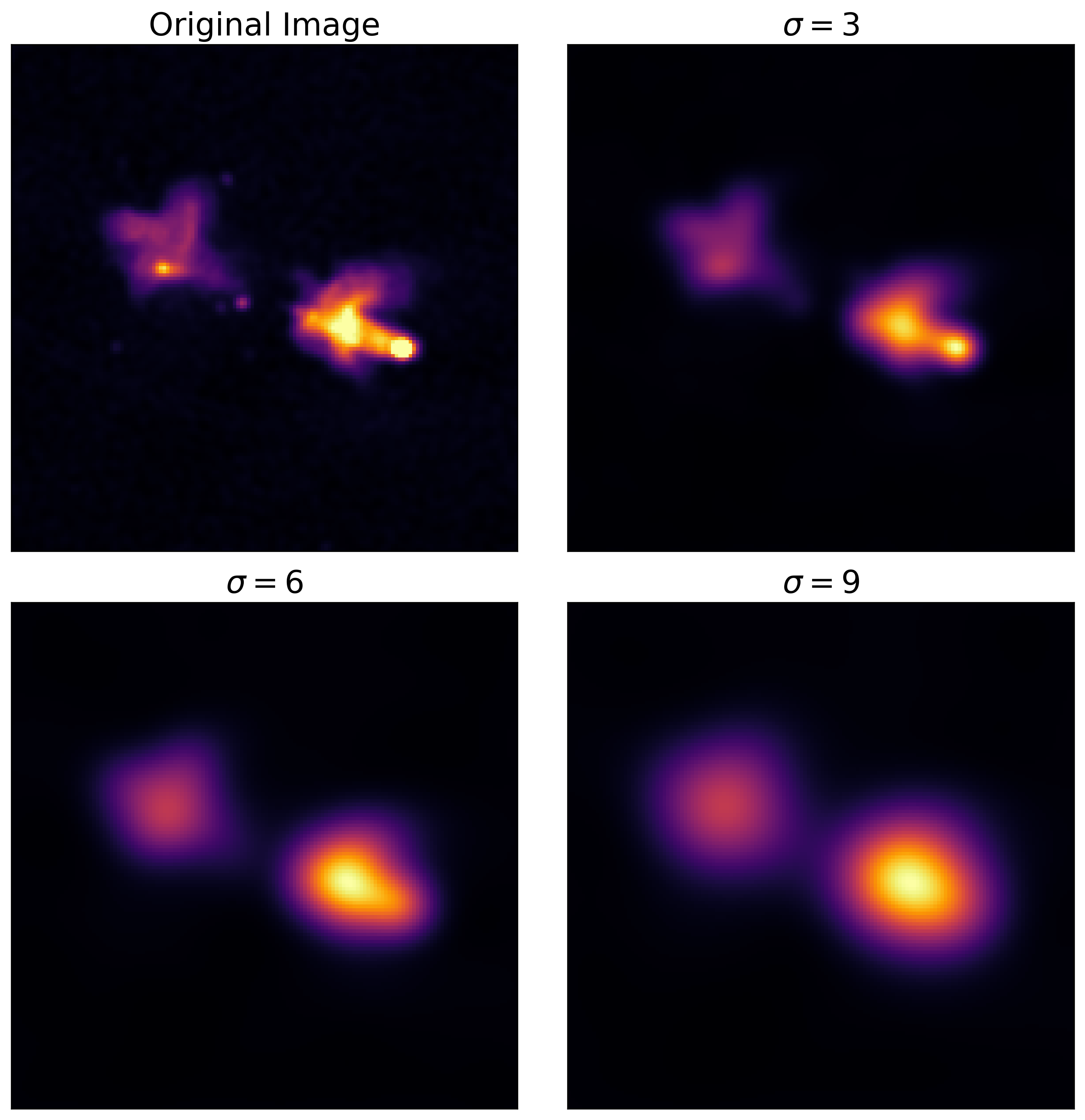}
    \caption{Effect of convolving radio image of 4C73.08 (top left) with 2D Gaussian kernels of increasing $\sigma$ in units of pixels.}
    \label{conv_eg}
\end{figure}

It is well known that metrics used in quantifying shapes, such as those presented in this work, are likely to be sensitive to the resolution of the images analysed. The shape of unresolved or marginally resolved sources (i.e., components only one or two beams wide) will be determined by the shape of the telescope synthesised beam, which is usually an elliptical Gaussian \citep{2023PASA...40...28B}. To understand the effect resolution has on each of our metrics, we convolve high resolution images with $3$ different $2$D Gaussian kernels, each with increasing standard deviation ($\sigma =$ $3$, $6$, and $9$) to produce images with progressively worse resolution. The convolution is performed in real space using direct kernel application. As the $\sigma$ increases, we begin to lose more and more structure until the lobes become unresolved (Figure~\ref{conv_eg}). We chose to convolve the 3CRR images for convenience, but the following analysis should be independent of the input images. 

\begin{figure*}[h!]
    \centering
    \includegraphics[width=0.75\linewidth]{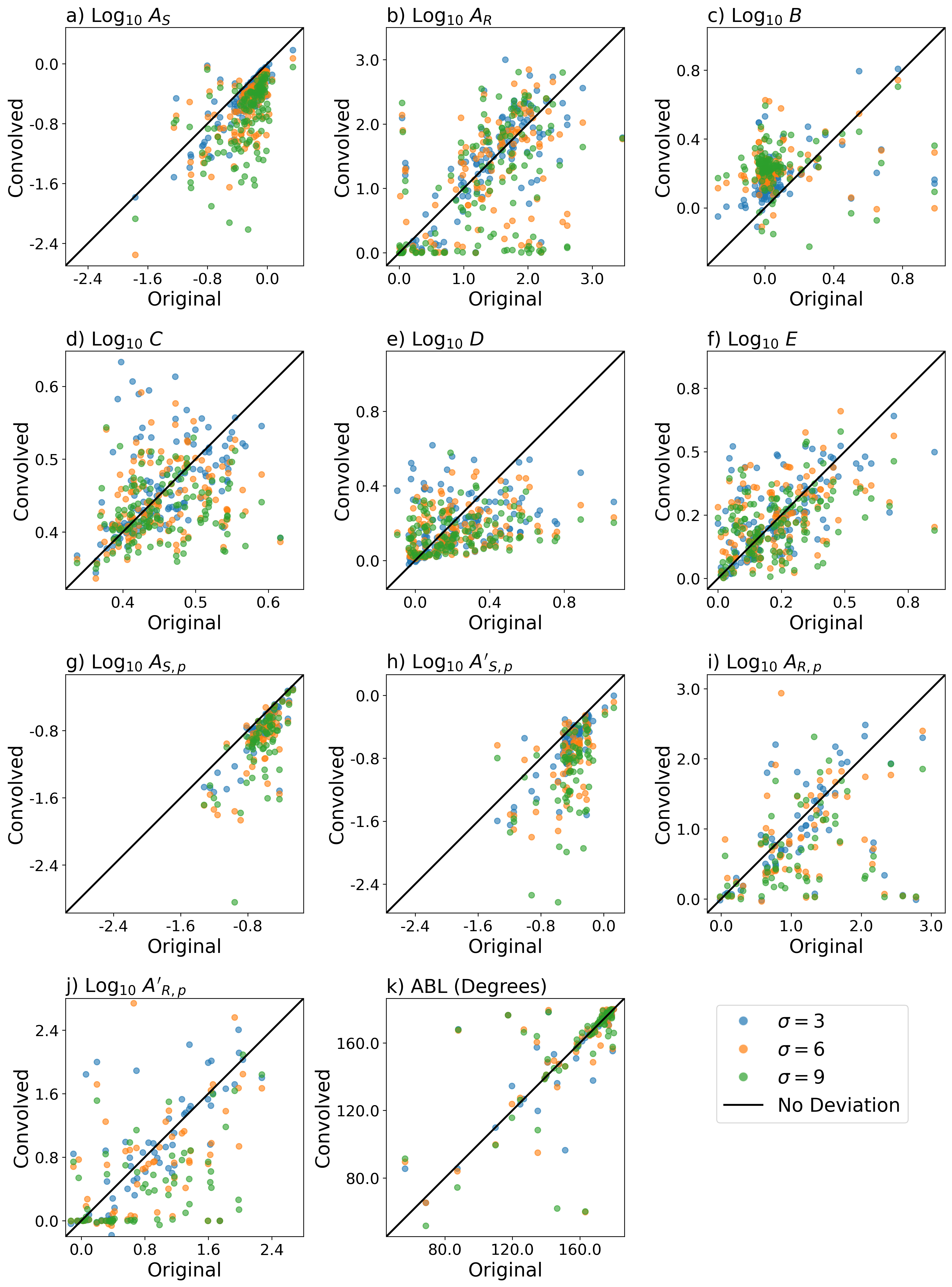}
    \caption{Comparison of the metric outputs of original 3CRR images and convolved versions. Blue, orange and green points correspond to images convolved with Gaussian kernels with $\sigma = 3$, $6$, and $9$ respectively. Black 1:1 line indicates no deviation in the metrics after convolving. Panels \textbf{a)-e)} have two points per source, one for each lobe. }
    \label{conv_comp_fig}
\end{figure*}

We apply each metric to the convolved images and compare them with the original values in Figure~\ref{conv_comp_fig}. We can see that the ABL calculation (Figure~\ref{conv_comp_fig}k) is the most robust against resolution differences. The scattered points in this metric result from location of the brightest pixel changing, often moving closer to the centre of a lobe after convolving. We see that convolving tends to produce smaller values for shape asymmetry measures ($A_S$, $A_{S,p}$ and $A'_{S,p}$). This is an expected result, as the flux in the lobes becomes more evenly distributed and circular in shape, resulting in lower asymmetry. This appears in Figure~\ref{conv_comp_fig}c, g, and h, where we can see smaller asymmetry values with increasing $\sigma$.

The other metrics show less intuitive trends. The process of convolving reduces the brightness of the peak pixel, spreading that flux over the surrounding pixels. For the metrics that use threshold masking (all apart from the ABL and asymmetry metrics), using $10\%$ of a lower peak brightness means more pixels lie above the threshold. With $B$ for example, more dim pixels around the edge of the lobe are retained. These pixels would produce a smaller average outer gradient making the overall ratio a larger number, moving the points above the $1:1$ line (Figure~\ref{conv_comp_fig}a). This effect would be increased as sigma gets larger, which is why the $\sigma=9$ points sit higher than the $\sigma=6$ and $\sigma=3$ points. For concentration, this may increase $r_{80}$ at a faster rate than the $r_{20}$, making the overall concentration measurement higher. 

We see for disorder and elongation that the convolved images produce fairly stable values, that then decrease gradually with increasing $\sigma$. This again is an expected result as the shape of the lobe becomes more circular with convolving, therefore making both disorder and elongation calculations approach unity. For disorder, the increase in the number of pixels will increase both the perimeter and area, however the perimeter$^2$ dependence will cause the disorder calculation to become much larger. For elongation, we expect the value to become closer to unity after convolving. It is possible that the increase in the number of dim pixels included some of the jet structure that was previously too dim, therefore increasing the overall elongation value. We have similar expectations with $A_R$, in that as we convolve the distribution of flux is smoothed, so $A_R$ should approach unity. As before with elongation, the inclusion of jet structure that was previously too dim, before convolving, can again increase the values of $A_R$.

The above analysis shows that the metrics can be highly resolution dependent, as evidenced in Figure~\ref{conv_comp_fig}, with many metrics showing high scatter when comparing original and convolved values. Higher resolution will enable the metrics to be more robust, as there is more structural information available to be extracted.

\section{Results and Analysis}

In this section, we present the results of applying the $ABCDE$, and source metrics to both the EMU-PS and 3CRR datasets to characterise and quantify the structure of the radio lobes. In Tables~\ref{tab:emu_data} and \ref{tab:3C_data}, we include a summary of the computed metric values for EMU and 3CRR sources, respectively. This data is available upon request. The name, host position, flux, and angular size for the EMU sources were obtained directly from RadioGalaxyNET \citep{gupta_radiogalaxynet_2024}. The name, flux, angular size, redshift and FR classification for the 3CRR sources were obtained from the Atlas of DRAGNs site \citep{leahy_atlas_2013}, with the host coordinates cross-matched from the Vizier catalogue of 3CRR sources \citep{laing2003_3crr_cat}. Note, all metric values are expressed in log$_{10}$, except for ABL. These results form the basis for the subsequent analyses, where we explore correlations between metrics and if our metrics can effectively characterise the structure and morphology of radio galaxies.

\begin{sidewaystable}
\scriptsize
\centering
    \begin{tabularx}{0.95\linewidth}{c c c c c c c c c c c c c c c c c c c c c c}
    \hline
    \hline
    \textbf{Name} & \textbf{RA} & \textbf{Dec} & \textbf{Flux} & \textbf{Ang Size} & \textbf{ABL} & \textbf{$A_{S1}$} & \textbf{$A_{R1}$} & \textbf{$A_{S2}$} & \textbf{$A_{R2}$} & \textbf{$A_{Sp}$} & \textbf{$A'_{Sp}$} & \textbf{$A_{Rp}$} & \textbf{$A'_{Rp}$} & \textbf{$B1$} & \textbf{$B2$} & \textbf{$C1$} & \textbf{$C2$} & \textbf{$D1$} & \textbf{$D2$} & \textbf{$E1$} & \textbf{$E2$} \\
    & \textbf{($^\circ$)} &  \textbf{($^\circ$)} & \textbf{(Jy)} & \textbf{(")} & \textbf{($\circ$)} & & & & & & & & & & & & & & & & \\
    \hline
    J200753.8$-$502330 & 301.98 & $-$50.39 & 0.00 & 109.58 & 168.14 & $-$0.35 & 1.40 & $-$0.50 & 0.78 & $-$0.50 & $-$0.41 & 1.22 & 0.70 & 0.04 & 0.03 & 0.45 & 0.43 & 0.71 & 0.32 & 0.22 & 0.45 \\
    J200825.1$-$525526 & 302.11 & $-$52.92 & 0.00 & 153.56 & 176.56 & $-$0.56 & 0.63 & $-$0.50 & 0.76 & $-$0.69 & $-$0.87 & 0.77 & 0.18 & $-$0.02 & 0.03 & 0.41 & 0.41 & 0.21 & 0.24 & 0.39 & 0.28 \\
    J200828.6$-$530013 & 302.12 & $-$53.00 & 0.08 & 104.78 & 162.74 & $-$0.60 & 1.00 & $-$0.69 & 0.59 & $-$1.27 & $-$0.97 & 0.41 & 0.17 & 0.04 & 0.03 & 0.39 & 0.37 & 0.11 & 0.08 & 0.34 & 0.28 \\
    J200920.8$-$494020 & 302.33 & $-$49.67 & 0.10 & 107.25 & 159.24 & $-$0.59 & 0.58 & $-$0.65 & 0.72 & $-$0.82 & $-$0.72 & 0.32 & 0.08 & $-$0.03 & 0.00 & 0.40 & 0.40 & 0.04 & $-$0.00 & 0.25 & 0.14 \\
    J200947.6$-$592758 & 302.45 & $-$59.47 & 0.00 & 90.75 & 117.14 & $-$0.97 & 0.02 & $-$0.68 & 0.58 & $-$1.05 & $-$1.07 & 0.16 & 0.02 & NaN & NaN & 0.39 & 0.36 & 0.31 & 0.30 & 0.25 & 0.19 \\
    J201017.6$-$522855 & 302.57 & $-$52.49 & 0.00 & 115.48 & 173.50 & $-$0.18 & 1.13 & $-$0.45 & 0.87 & $-$0.50 & $-$0.24 & 1.09 & 0.94 & 0.11 & $-$0.05 & 0.44 & 0.45 & 0.41 & 0.01 & 0.22 & 0.13 \\
    J201051.4$-$525836 & 302.72 & $-$52.98 & 0.03 & 137.84 & 165.68 & $-$0.51 & 0.85 & $-$0.58 & 0.59 & $-$0.80 & $-$0.52 & 0.41 & 0.32 & -0.00 & 0.04 & 0.44 & 0.39 & 0.31 & 0.01 & 0.47 & 0.09 \\
    J201052.7$-$582128 & 302.72 & $-$58.36 & 0.01 & 105.75 & 164.57 & $-$0.36 & 0.82 & $-$0.22 & 0.57 & $-$0.59 & $-$0.47 & 0.79 & 0.80 & 0.04 & 0.06 & 0.46 & 0.45 & 0.13 & 0.55 & 0.20 & 0.02 \\
    J201055.3$-$492051 & 302.73 & $-$49.34 & 0.02 & 158.69 & 162.90 & $-$0.57 & 0.68 & $-$0.54 & 0.32 & $-$1.15 & $-$0.94 & 0.03 & $-$0.04 & 0.03 & 0.01 & 0.41 & 0.43 & 0.01 & 0.03 & 0.14 & 0.17 \\
    J201107.5$-$525706 & 302.78 & $-$52.95 & 0.03 & 82.40 & 178.51 & $-$0.63 & 1.22 & $-$1.01 & 0.01 & $-$0.93 & $-$1.01 & 0.20 & $-$0.01 & 0.22 & $-$0.06 & 0.41 & 0.43 & 0.16 & 0.06 & 0.19 & 0.12 \\
    \hline
    \hline
    \end{tabularx}
    \caption{Top $10$ rows of $ABCDE$, and source metric calculations for the EMU-PS sources.}
    \label{tab:emu_data}
\end{sidewaystable}

\begin{sidewaystable}
\centering
\scriptsize
    \begin{tabularx}{0.95\linewidth}{c c c c c c c c c c c c c c c c c c c c c c c c c c}
    \hline
    \hline
    \textbf{Name} & \textbf{RA} & \textbf{Dec} & \textbf{Flux} & \textbf{Ang Size} & \textbf{ABL} & \textbf{$A_{S1}$} & \textbf{$A_{R1}$} & \textbf{$A_{S2}$} & \textbf{$A_{R2}$} & \textbf{$A_{Sp}$} & \textbf{$A'_{Sp}$} & \textbf{$A_{Rp}$} & \textbf{$A'_{Rp}$} & \textbf{$B1$} & \textbf{$B2$} & \textbf{$C1$} & \textbf{$C2$} & \textbf{$D1$} & \textbf{$D2$} & \textbf{$E1$} & \textbf{$E2$} & \textbf{z} & \textbf{FR} \\
    & \textbf{($^\circ$)} &  \textbf{($^\circ$)} & \textbf{(Jy)} & \textbf{(")} & \textbf{($^\circ$)} & & & & & & & & & & & & & & & & & & \textbf{Class}\\
    \hline
    4C12.03 & 2.47 & 12.73 & 10.90 & 232.00 & 179.36 & $-$0.04 & 1.69 & $-$0.71 & 0.11 & $-$0.55 & $-$0.38 & 0.58 & 1.98 & $-$0.12 & 0.68 & 0.43 & 0.44 & 0.29 & 0.57 & 0.41 & 0.23 & 0.16 & II \\
    3C19 & 10.23 & 33.17 & 13.20 & 6.80 & 173.08 & $-$0.07 & 1.40 & $-$0.08 & 1.30 & $-$0.58 & $-$0.31 & 1.13 & 0.91 & 0.02 & 0.08 & 0.42 & 0.49 & 0.03 & 0.22 & 0.10 & 0.15 & 0.48 & II \\
    3C20 & 10.79 & 52.06 & 46.80 & 53.10 & 174.63 & $-$0.07 & 1.88 & $-$0.14 & 1.95 & $-$0.70 & $-$0.43 & 0.97 & 0.80 & 0.06 & 0.07 & 0.47 & 0.42 & 0.34 & 0.01 & 0.08 & 0.13 & 0.17 & II \\
    3C28 & 13.96 & 26.41 & 17.80 & 45.60 & 157.69 & $-$0.44 & 1.47 & $-$0.12 & 2.18 & $-$0.62 & $-$0.31 & 2.33 & 0.67 & $-$0.11 & $-$0.14 & 0.41 & 0.40 & 0.30 & 0.21 & 0.02 & 0.21 & 0.20 & II \\
    3C31 & 16.85 & 32.41 & 18.30 & 2700.00 & 87.27 & $-$0.17 & 2.02 & $-$0.80 & 0.07 & $-$0.46 & $-$0.64 & 0.72 & 0.25 & 0.04 & NaN & 0.47 & 0.42 & 0.49 & 0.21 & 0.58 & 0.10 & 0.02 & I \\
    3C33 & 17.22 & 13.34 & 59.30 & 257.00 & 177.31 & $-$0.44 & 1.67 & $-$0.20 & 2.08 & $-$0.54 & $-$0.43 & 0.78 & 0.34 & $-$0.04 & $-$0.01 & 0.41 & 0.42 & 0.02 & 0.04 & 0.07 & 0.28 & 0.06 & II \\
    3C33.1 & 17.43 & 73.20 & 14.20 & 238.70 & 172.26 & $-$0.13 & 2.39 & $-$0.09 & 2.54 & $-$0.59 & $-$0.45 & 1.09 & 1.93 & 0.02 & 0.09 & 0.46 & 0.48 & 0.28 & 0.66 & 0.26 & 0.37 & 0.18 & II \\
    3C42 & 22.13 & 29.05 & 13.10 & 31.00 & 165.69 & $-$0.15 & 2.24 & $-$0.31 & 1.65 & $-$0.74 & $-$0.43 & 0.75 & 0.62 & 0.03 & -0.02 & 0.43 & 0.41 & 0.02 & $-$0.01 & 0.14 & 0.05 & 0.40 & II \\
    3C46 & 23.87 & 37.90 & 11.10 & 164.00 & 173.00 & $-$0.13 & 1.74 & $-$0.15 & 1.35 & $-$0.51 & $-$0.19 & 1.69 & 1.60 & $-$0.01 & 0.09 & 0.48 & 0.49 & 0.18 & 0.20 & 0.41 & 0.32 & 0.44 & II \\
    3C47 & 24.10 & 20.96 & 28.80 & 77.50 & 178.38 & $-$0.13 & 2.32 & $-$0.29 & 2.03 & $-$0.70 & $-$0.44 & 0.63 & 0.42 & NaN & 0.02 & 0.42 & 0.46 & 0.19 & 0.13 & 0.39 & 0.28 & 0.42 & II \\
    \hline
    \hline
    \end{tabularx}
    \caption{Top $10$ rows of $ABCDE$, and source metric calculations for the 3CRR sources.}
    \label{tab:3C_data}
\end{sidewaystable}

\subsection{Single-Lobe Metrics}

\begin{sidewaysfigure*}[htp]
    \centering
    \includegraphics[width=0.9\linewidth]{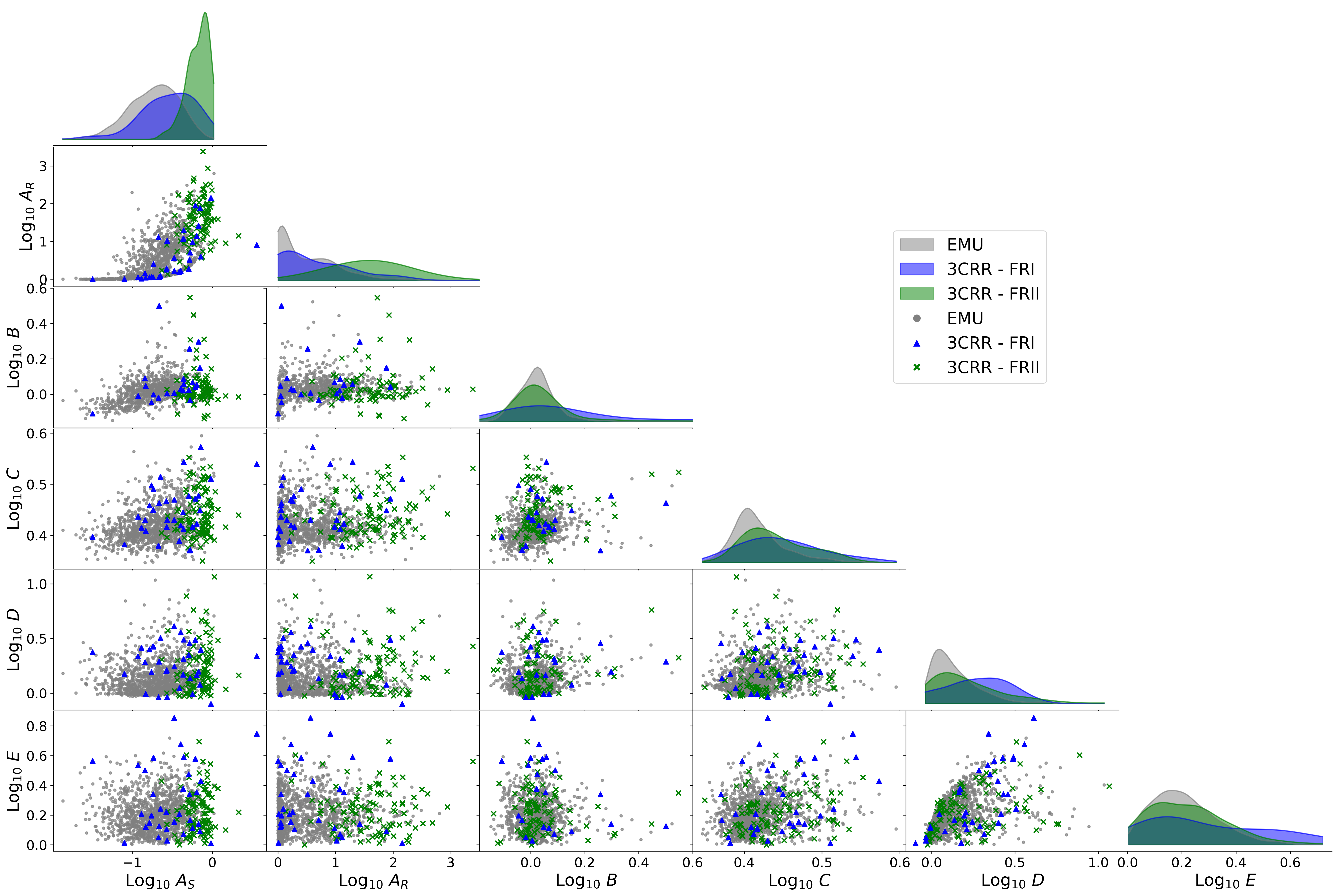}
    \caption{Pairwise comparisons of the single-lobe ($ABCDE$) metrics calculated on radio galaxies from the EMU-PS and 3CRR surveys. The scatter plots show the relationships between these metrics, with normalised kernel density estimates (KDEs) along the diagonal displaying the distribution of each. EMU sources are represented by grey dots, and 3CRR sources colour-coded by their FR classification: FRI sources are shown with blue triangles and FRII sources with green crosses. All panels here have two points per source, one for each lobe.}
    \label{lobe-lobe_metrics_fig}
\end{sidewaysfigure*}

Figure~\ref{lobe-lobe_metrics_fig} shows pairwise combinations of the single-lobe calculations ($ABCDE$), with the normalised kernel density estimate (KDE) for each metric in the diagonals. For most of the metrics, the EMU and 3CRR sources occupy similar areas of parameter space, indicating the minimum image area of $20$ ASKAP beams for the EMU sources allows them to be sufficiently resolved for the metric results to be comparable with the 3CRR data. While some metric combinations show slight correlations (e.g., $B$ and $C$) or more complicated relationships (e.g., $A_S$ and $A_R$) we do see high scatter in each subplot, indicating that these metrics my be independent. We calculated the Spearman rank correlation coefficient \citep[$\rho$,][]{siegel_nonparametric_1988, conover1999practical} on each of the unique metric combinations from Figure~\ref{lobe-lobe_metrics_fig}. Values of $\rho \approx 0$ correspond to little to no correlation, with $\rho \approx \pm 1$ correspond to a strong positive or negative correlation, respectively. We found that most of the metric combinations had $-0.054 < \rho < 0.27$, with the exception of $A_S$-$A_R$ ($\rho$ = $0.72$) and $D$-$E$ ($\rho = 0.57$). This indicates that the metrics are, in fact, largely independent of each other and therefore measure fundamentally different structural features of a radio galaxy lobe. 

Since the FRI/II dichotomy has been a very productive scheme for uncovering the physics of radio galaxies, we will explore to what extent these metrics can reproduce this dichotomy. Both FR-types occupy similar regions in metric space for the majority of our metrics, with the exception $A_S$, $A_R$ where we see stronger separation between FRI and FRII sources. FRI sources also tend to have a broad range of values, when compared to the FRII sources, particularly in the $B$, $C$, and $E$ metrics. This reinforces the need for additional metrics to completely characterise the shape of radio lobes. We explore correlations between metrics and FR distributions in more detail in subsequent sections.

\subsubsection{Asymmetry}\label{Asymm_analysis}

\begin{figure}[h!]
    \centering
    \includegraphics[width=\linewidth]{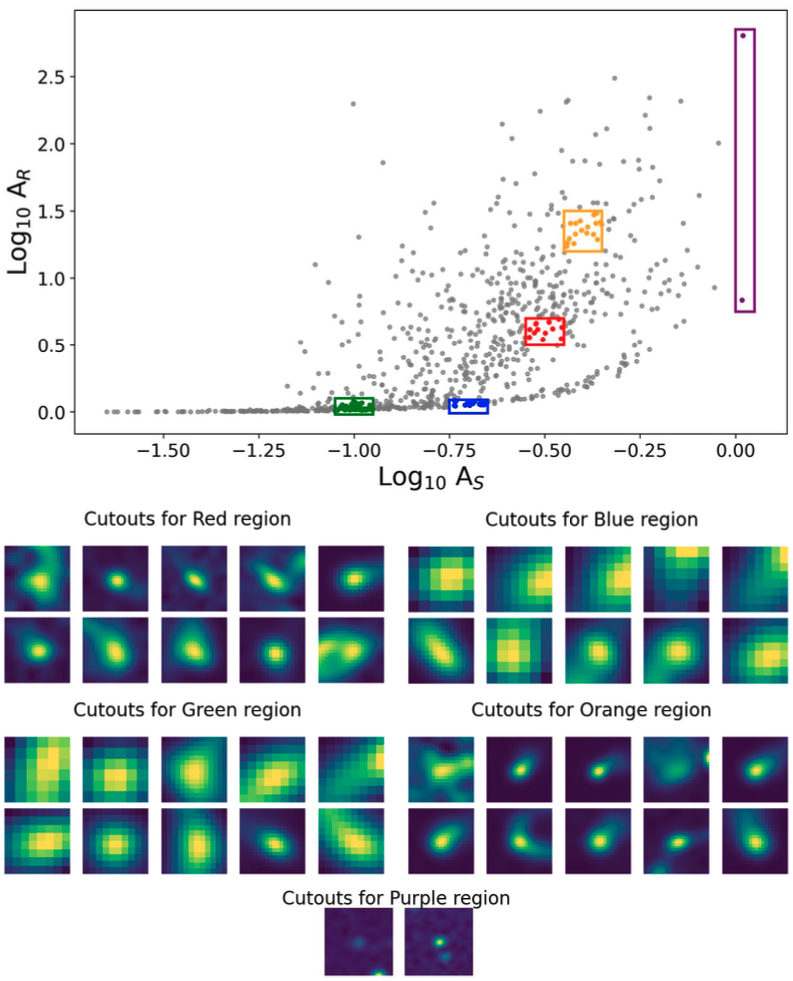}
    \caption{Enlarged version of the $A_S$-$A_R$ sub-panel from Figure~\ref{lobe-lobe_metrics_fig} showing only EMU data. In this figure different regions of this parameter space are highlighted by different colours. Panels below are lobe cutouts for each highlighted region.}
    \label{asym_regions}
\end{figure}

Evident in both the scatter and KDE panels in Figure~\ref{lobe-lobe_metrics_fig}, the FRI and FRII sources are found in different areas of the asymmetry parameter space. We see that for both $A_S$ and $A_R$ that the FRI sources tend be less asymmetric than FRII sources. While the EMU sources share the same region of parameter space as the FRIs there is evidence of a bimodal distribution, highlighted in the KDE of the $A_R$ metric. There is a highly concentrated population at very low $A_R$, and another more scattered population at moderate to high $A_R$. The initial comparison of EMU and 3CRR sources indicates that the low $A_R$ values coincide with the FRIs and the moderate to high $A_R$ values with FRIIs. We explore the $A_S$-$A_R$ parameter space further in Figure~\ref{asym_regions}. 

The concentrated tail at low $A_R$ values (e.g., the blue and green regions) may be produced when the brightest pixel is near the edge of the lobe cutout, therefore limiting the size of the cutout and causing the overall flux distribution to appear more uniform. This will result in lower values for the $A_R$ metric, forming the tail with $\log_{10}(A_R)$ values close to $0$. We can relate this to the abundance of FRI sources present in the tail. FRI sources are often referred to as core-brightened, with the peak brightness occurring in the jets near the core. The code splits an image of a radio galaxy in half under the assumption that there will be a radio lobe in each half. As the FRI hotspots tend to be closer to the core, the brightest pixel for an FRI source would therefore be near the border of the lobe cutout, limiting the number of pixels in the asymmetry calculations. The tail is therefore likely an artefact of both the flux distribution of the sources and of how we isolate a lobe in our algorithm. For objects found in the more scattered section of this parameter space (e.g., the red and orange regions), the asymmetry metric appears to be working as expected, with the scatter resulting from actual variance in lobe structure in different sources. 

In the purple region, there are two measurements with $\log_{10}(A_S)>0$, which should not occur for a normally distributed source. We can see in the lobe cutouts for this region of parameter space that both contain two separate circular sources, with one near the centre and the other offset. From visual inspection, the central sources appear to have low rotational asymmetry, and will not contribute much to the asymmetry value. For the off-centre sources however, the rotation by $180^{\circ}$ will produce two bright spots after taking the absolute value, corresponding to a high $A_S$ value. We should also note that both of these sources have relatively low overall brightness, and, we can see some of the noise fluctuations in surrounding pixels. 

We conducted simulations to investigate the features observed in Figure~\ref{asym_regions}, particularly what produces the concentrated tail and what causes sources to have $\log_{10}(A_S)>0$. To simulate a perfectly symmetric lobe (red points), with the only asymmetry coming from noise fluctuations we first populate a 1D array with values ranging between $1\times 10^{-5}$ and $1$, referred to as array $x$. We then generate a second 1D array, $y$, which is identical to array $x$ except for the addition of random noise, normally distributed around zero with an amplitude of $0.1$:
\begin{equation*}
    y = x + \text{noise}~,
\end{equation*}

\noindent where we then calculate $A_S$ and $A_R$ by substituting $x$ for lobe $1$ and $y$ for lobe $2$ into Equations~\ref{shape_res} and \ref{rat_res}, respectively.

\begin{figure}[h]
    \centering
    \includegraphics[width=\linewidth]{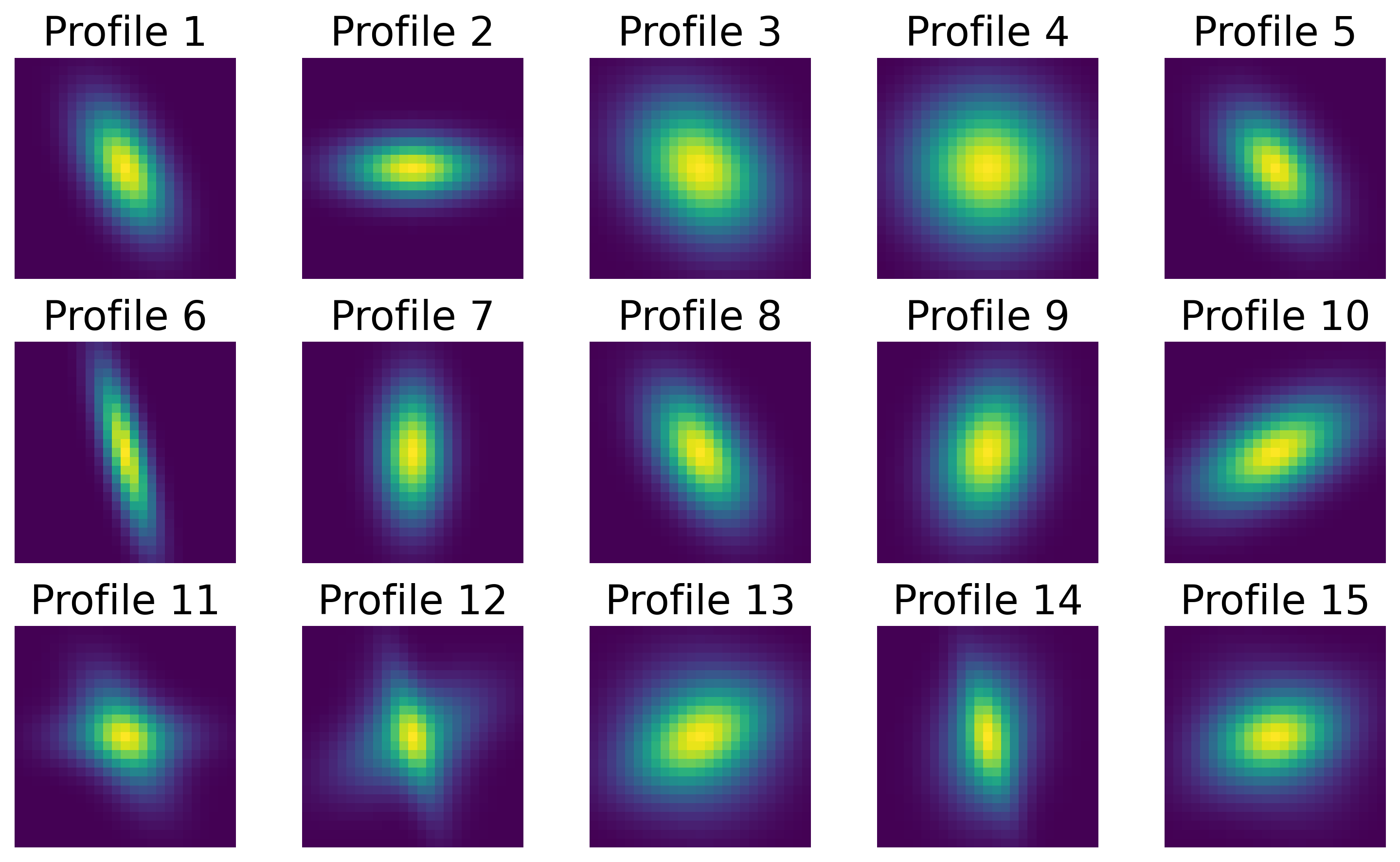}
    \caption{Collection of the $15$ different 2D Gaussian profiles used in the simulations.}
    \label{sim_gauss_prof}
\end{figure}

To simulate an asymmetric lobe (blue points), we generated $15$ different $25 \times 25$ Gaussian profiles, with ranging intensities, $\sigma$, skew, and rotation (Figure~\ref{sim_gauss_prof}). We loop over each of the profiles, systematically adding noise until we have the same number of data points as our EMU-PS data ($480$), repeating the asymmetry calculations. The noise added in this loop is normally distributed around zero, with amplitude ranging from $0.015$ to $0.9$.

Finally for the noise-only simulation (green points), we randomly populate a $15 \times 15$ array with Gaussian noise matching the noise distribution of the EMU-PS, i.e., centred on zero with $\sigma = 23\,\mu$Jy/beam \citep{norris_evolutionary_2021}.

\begin{figure}[h!]
    \centering
    \includegraphics[width=\linewidth]{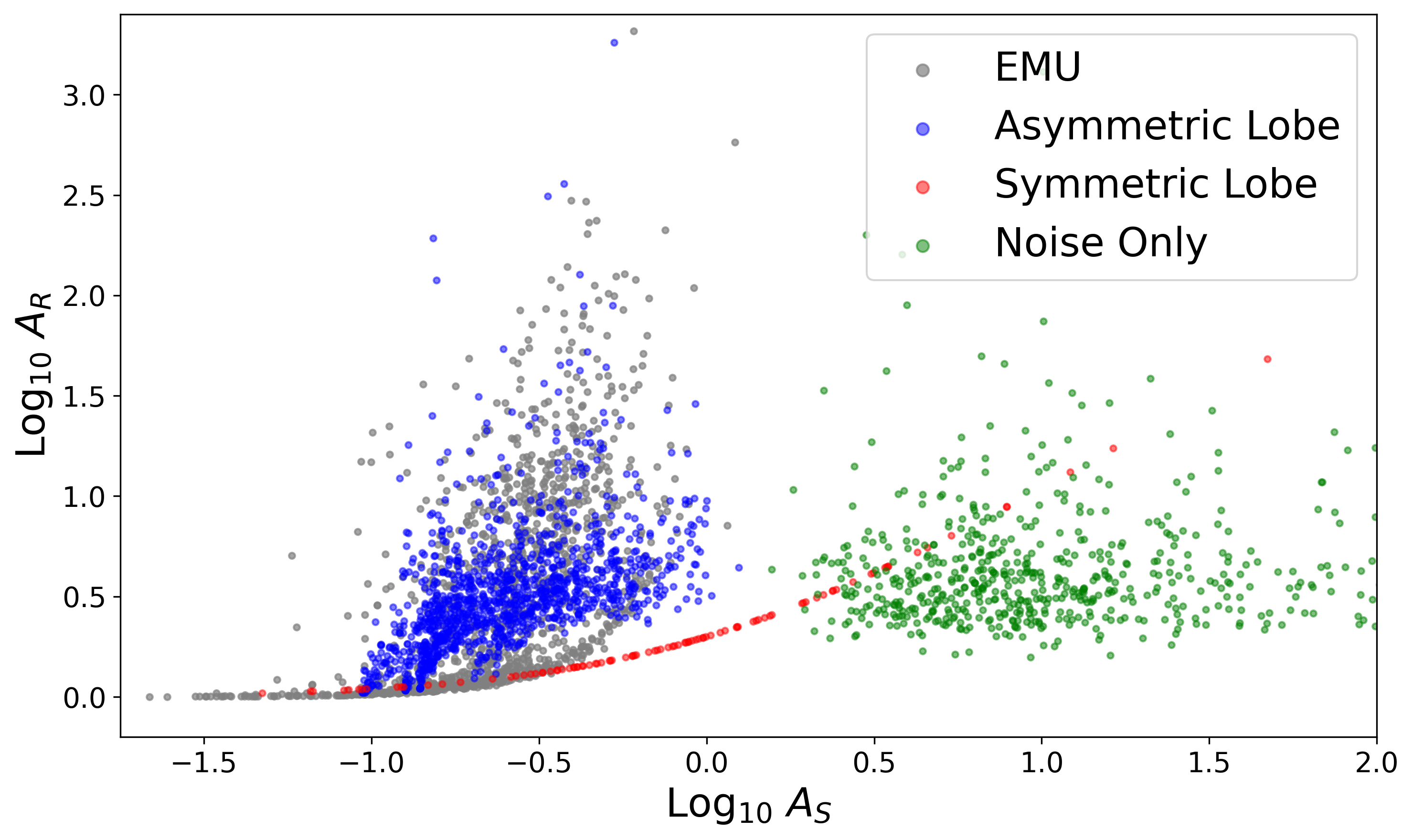}
    \caption{Enlarged version of the $A_S$-$A_R$ sub-panel from Figure~\ref{lobe-lobe_metrics_fig} with only EMU data, comparing how the asymmetry parameters characterise different simulated data. Grey, blue, red, and green correspond to data points from the EMU-PS, simulated asymmetric lobes, simulated symmetric lobes, and Gaussian distributed noise, respectively.}
    \label{asym_sims}
\end{figure}

The asymmetry calculations on the simulated data (Figure~\ref{asym_sims}) revealed that the noise-only points explicitly exist at $\log_{10}(A_S)>0$. This is likely due to normalising by the total flux. The sum of an array populated by Gaussian noise distributed around zero approaches $0$ as we increase the number of points in the array. This means that for the noise-only points (i.e., sources with poor signal-to-noise), we normalise by a very small number, leading to $\log_{10}(A_S)$ values greater than $0$. In this work we wanted to investigate how the metrics perform on a variety of different sources, and we do not include any minimum signal to noise cutoff. 

While the initial asymmetry results (both $A_S$ and $A_R$) taken at face value suggest that FRII sources tend to be more asymmetric than FRIs, they need to be interpreted with caution. The way $A_S$ and $A_R$ are defined, using the brightest pixel as a centre of rotation, the tendency is for FRI sources to have fewer pixels, and a more symmetrical structure within the cutout for each lobe. As discussed earlier, a limited number of pixels (i.e., if the bright spots of a lobe are near the edge of a lobe cutout) can produce a low asymmetry value. Conversely, FRII lobes tend to have hotspots, and hence a brightest pixel, positioned such that more of the actual lobe structure is included in the measurement, leading to higher values of asymmetry. This is not to say that these measures are wrong, but it is important to interpret them given how they are defined, reflecting the degree of asymmetry only within the region sampled.

\subsubsection{Blurriness and Concentration}

\begin{figure}[h!]
    \centering
    \includegraphics[width=\linewidth]{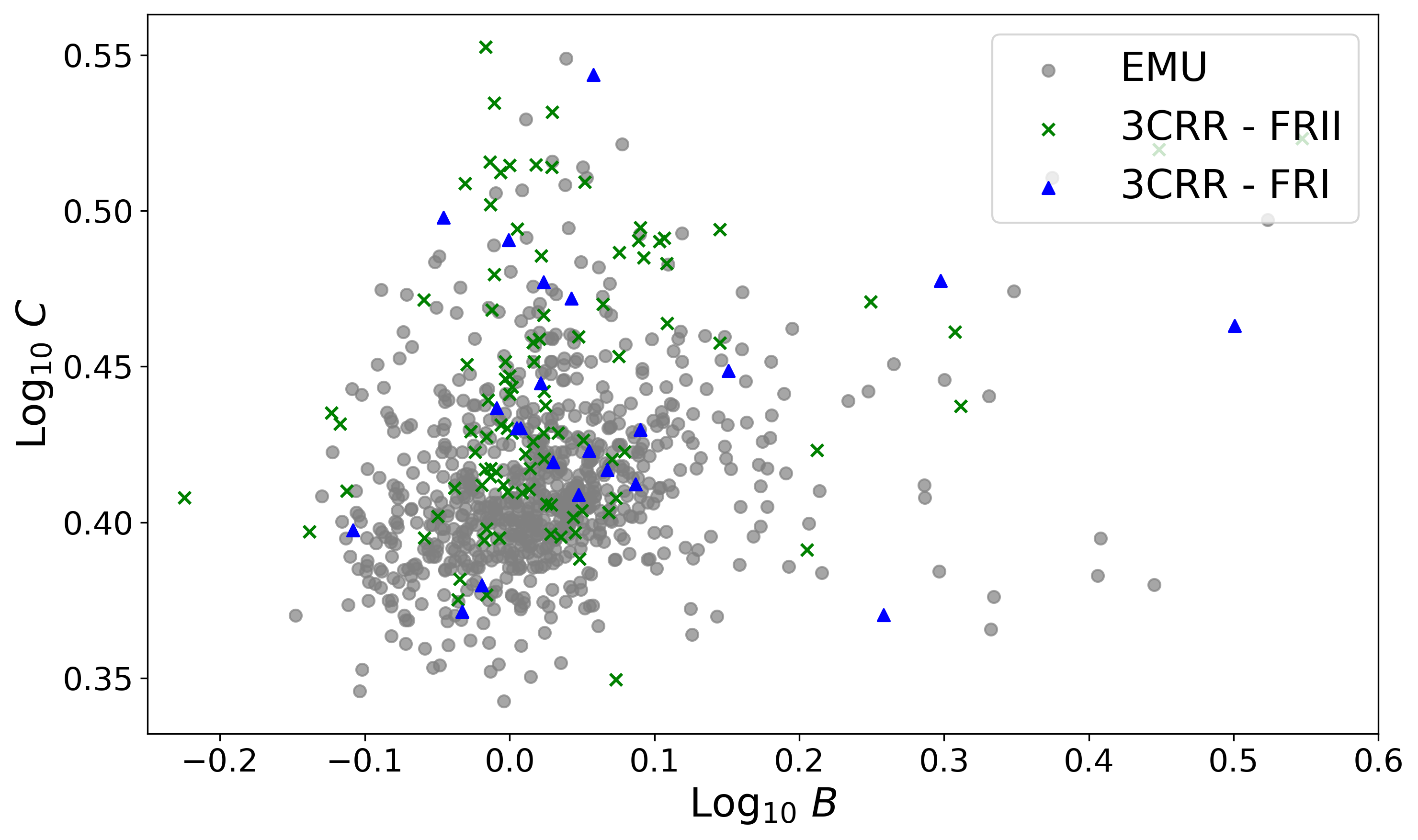}
    \caption{Enlarged version of the $B$-$C$ sub-panel from Figure~\ref{lobe-lobe_metrics_fig}. EMU sources are represented by grey dots, and 3CRR sources colour-coded by their FR classification: FRI sources are shown with blue triangles and FRII sources with green crosses.}
    \label{g_diff_conc}
\end{figure}

Figure~\ref{g_diff_conc} reveals a weak positive correlation between the $B$ and $C$ metrics ($\rho = 0.27$) for the EMU and for the 3CRR data ($\rho = 0.24$). Small $B$ values can be indicative of an edge-brightened source, possibly identifying lobes that are compressed by the environment. Conversely larger $B$ values indicate a rapid flux drop from the brightest pixel to the outer parts of the lobe (i.e., higher concentration of flux), corresponding to a edge-dimmed source, that could exist in a lower density environment. These implications can be tested by investigating the number density of surrounding galaxies \citep[e.g.,][]{brough_galaxy_2013} for sources with different $B$ and $C$ values. We can also see the FRI and FRII sources span very similar ranges for both $B$ and $C$, (although FRIIs may have slightly higher concentration values than FRIs). When developing $B$ and $C$, it was anticipated that the hotspots would result in higher concentration (and blurriness) values for FRII sources, potentially showing higher separation between FRI and FRII types in this metric space. These findings further highlight the limitations of the binary FR classification. Quantitative metrics, such as $B$ and $C$, provide more nuanced descriptions of radio galaxy morphology. This reinforces the need to move away from the binary FR classification system and instead use quantitative descriptive metrics for a given source.

\subsubsection{Disorder and Elongation}

\begin{figure}[h!]
    \centering
    \includegraphics[width=\linewidth]{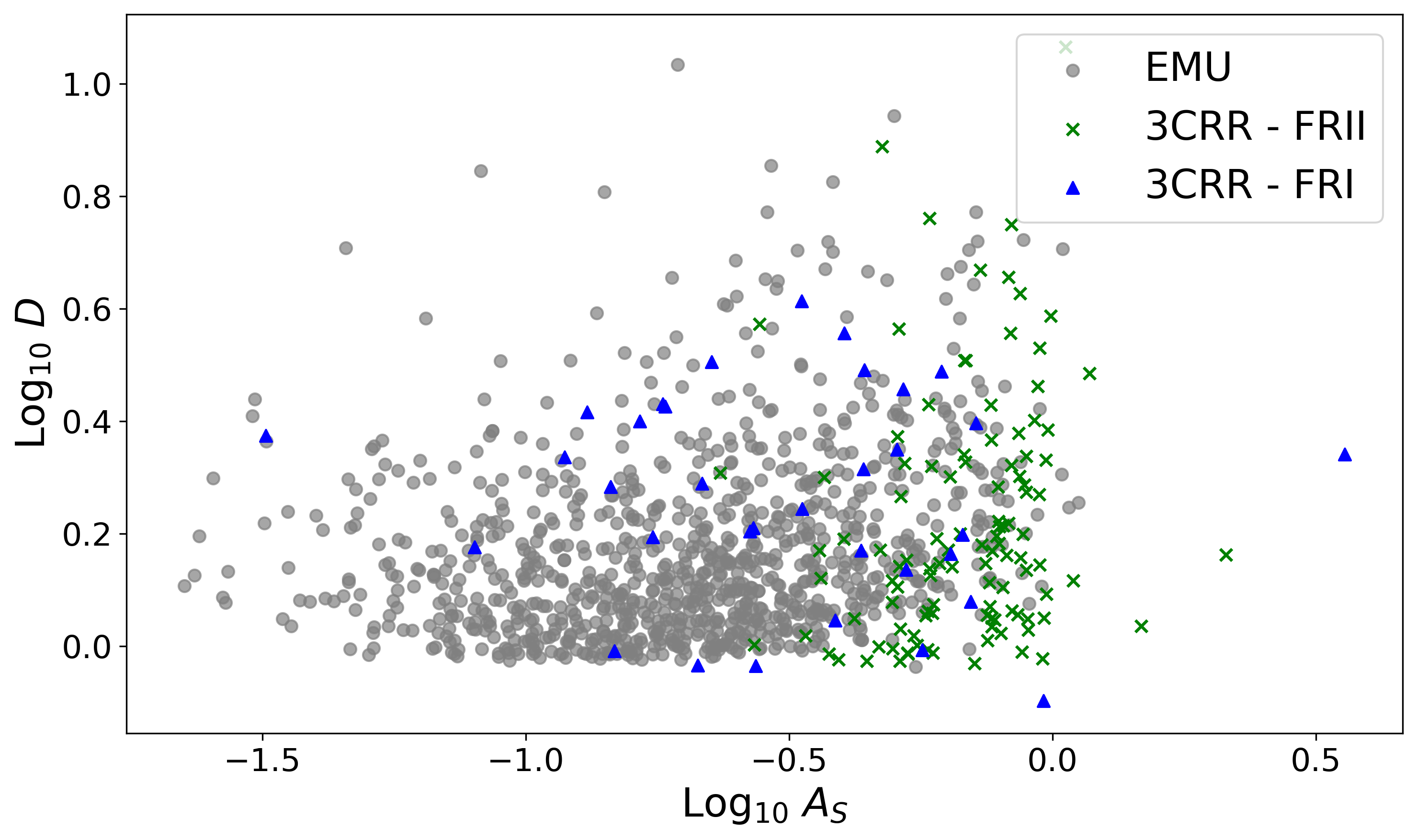}
    \caption{Enlarged version of the $A_S$-$D$ sub-panel from Figure~\ref{lobe-lobe_metrics_fig}. EMU sources are represented by grey dots, and 3CRR sources colour-coded by their FR classification: FRI sources are shown with blue triangles and FRII sources with green crosses.}
    \label{AS_D}
\end{figure}

\begin{figure}[h!]
    \centering
    \includegraphics[width=\linewidth]{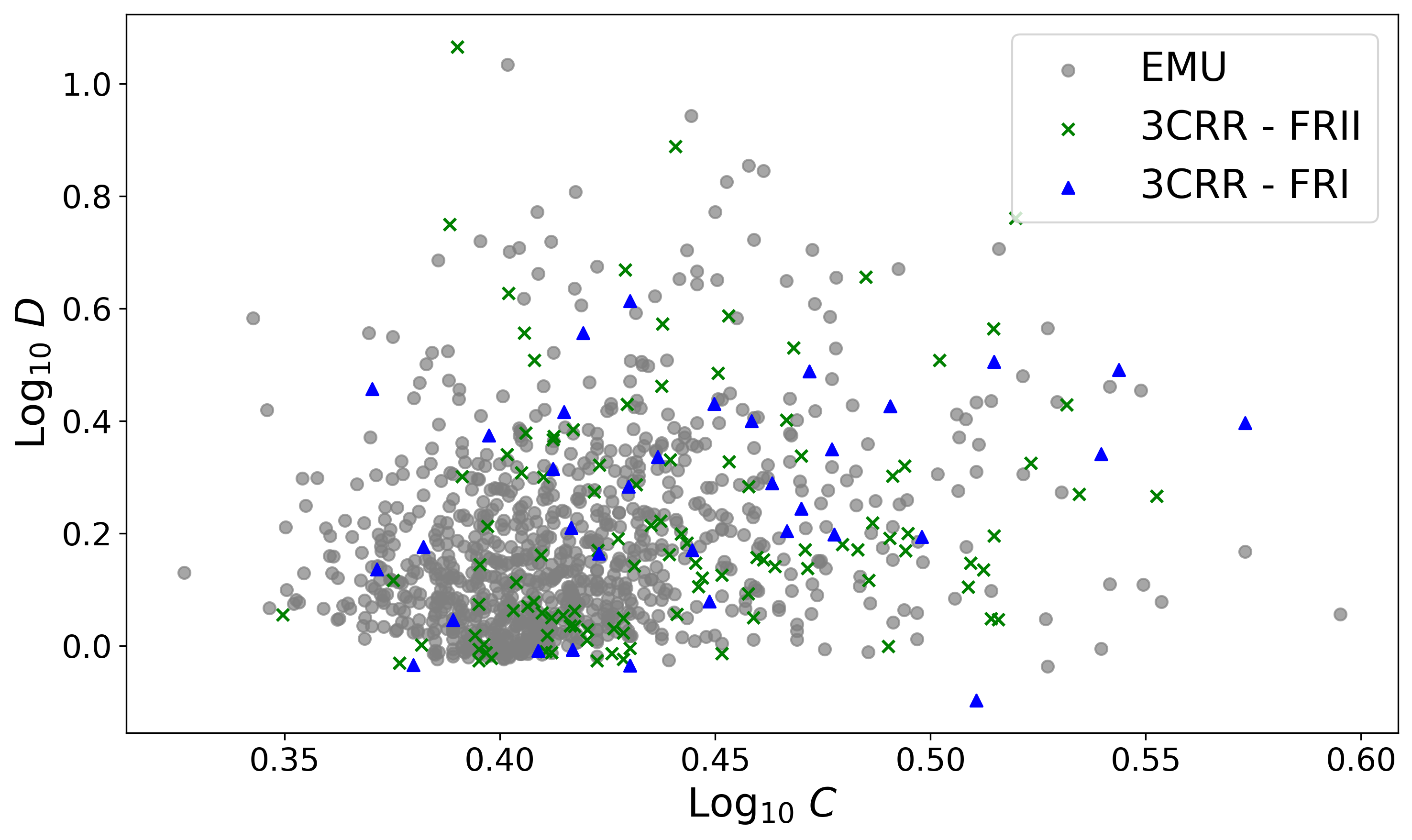}
    \caption{Enlarged version of the $C$-$D$ sub-panel from Figure~\ref{lobe-lobe_metrics_fig}. EMU sources are represented by grey dots, and 3CRR sources colour-coded by their FR classification: FRI sources are shown with blue triangles and FRII sources with green crosses.}
    \label{C_D}
\end{figure}

We explore connections between $A_S$-$D$ and $C$-$D$ in Figures~\ref{AS_D} and \ref{C_D}, respectively. We find that $A_S$ and $D$ are very weakly correlated ($\rho = 0.16$ and $0.077$ for EMU and 3CRR, respectively), and that there are no sources with both high disorder and low asymmetry. Similarly, we find that $C$ and $D$ are weakly correlated ($\rho = 0.27$ for both EMU and 3CRR), and there are no sources with both high disorder and low concentration. Highly disordered structures tend to coincide with more asymmetric or diffuse features, making it less likely for a source to simultaneously have low asymmetry or high concentration. From the KDE distributions in Figure~\ref{lobe-lobe_metrics_fig}, while the FRI sources appear to have slightly larger values of $D$ than FRII sources, this distinction is not clear. This could be due to the lower jet power and brightness of the FRI sources than those FRII sources. The lobe structure may therefore be more sensitive to noise fluctuations, increasing the overall disorder of a lobe. The lack of FR separation in $C$ and $D$ suggest that either these metrics may not be sufficiently sensitive to hotspots and jet power, or that other factors, such as variations in lobe brightness distribution and environmental effects, play a dominant role in determining concentration and disorder.

\begin{figure}[h!]
    \centering
    \includegraphics[width=\linewidth]{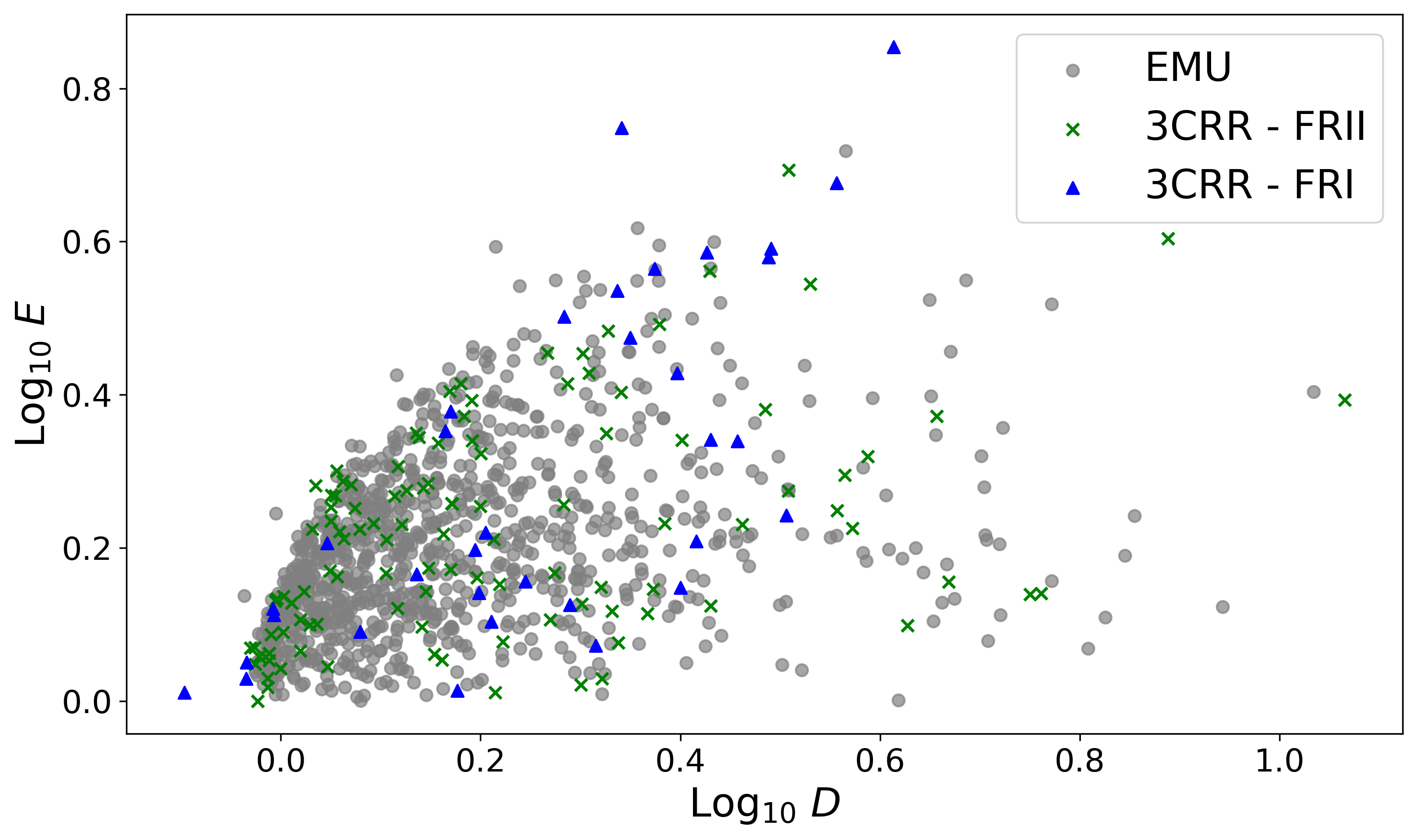}
    \caption{Enlarged version of the $D$-$E$ sub-panel from Figure~\ref{lobe-lobe_metrics_fig}. EMU sources are represented by grey dots, and 3CRR sources colour-coded by their FR classification: FRI sources are shown with blue triangles and FRII sources with green crosses.}
    \label{D_E_fig}
\end{figure}

We find $D$ and $E$ are moderately correlated ($\rho = 0.57$ and $0.51$ for EMU and 3CRR sources, respectively). The comparison between the $D$ and $E$ metrics in Figure~\ref{D_E_fig} highlights a general logarithmic envelope, with no sources exhibiting both high elongation and low disorder. Highly elongated jets or lobes will be more susceptible to environmental interactions, turbulence, and instabilities, all of which can contribute to increased disorder. There is also no clear separation between FR types in this parameter space. This suggests that disorder and elongation alone are not primary distinguishing factors in the classical FR classification. Instead, these metrics may be more reflective of environmental influences and source dynamics rather than intrinsic differences between FR types.

\subsection{Source Metrics}

\begin{figure*}[htp!]
    \centering
    \includegraphics[width=\linewidth]{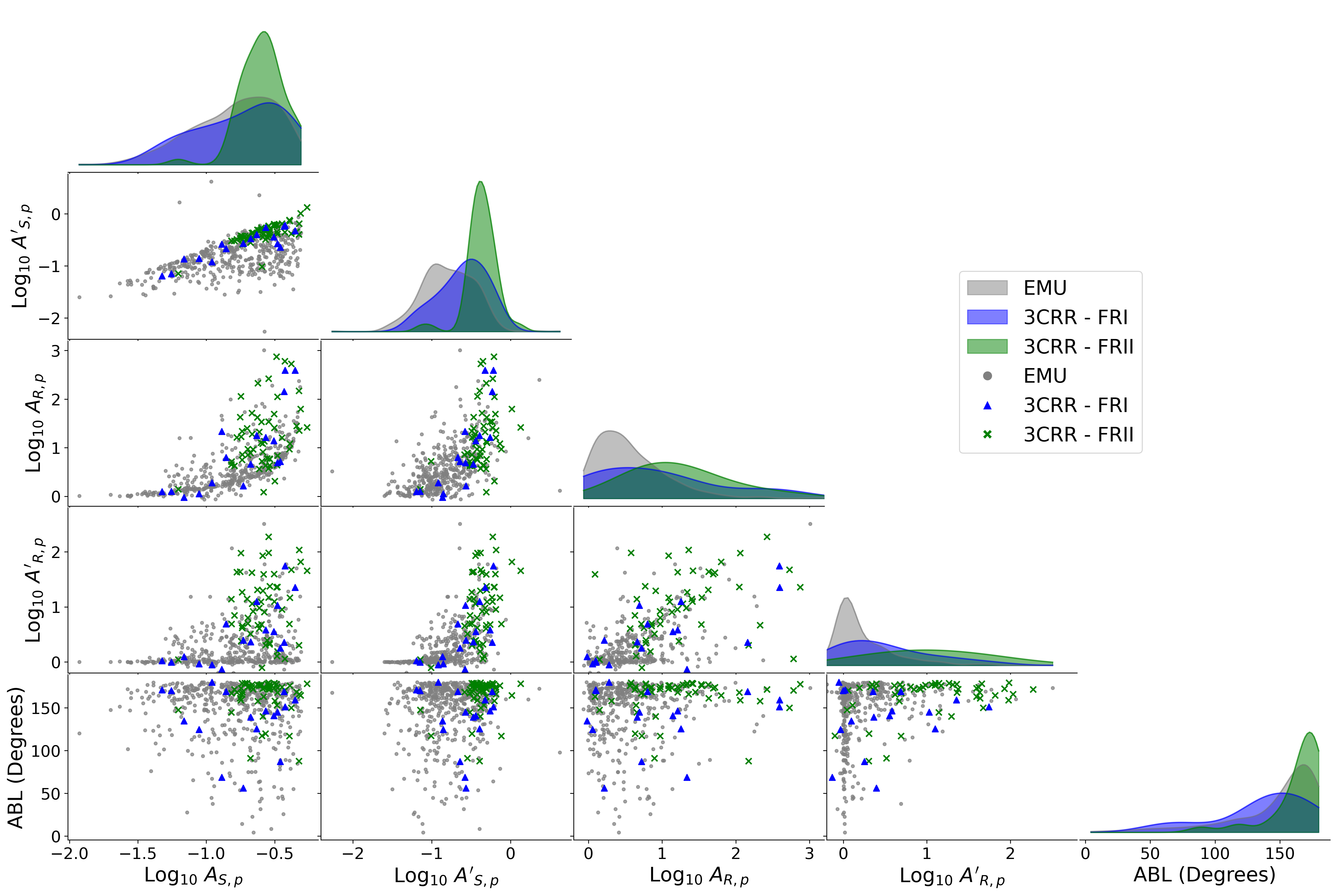}
    \caption{Pairwise comparisons of the source metrics (ABL, $A_{S,p}$, $A'_{S,p}$, $A_{R,p}$, $A'_{R,p}$) calculated on radio galaxies from the EMU-PS and 3CRR surveys. The scatter plots show the relationships between these metrics, with normalised KDEs along the diagonal displaying the distribution of each. EMU sources are represented by grey dots, and 3CRR sources colour-coded by their FR classification: FRI sources are shown with blue triangles and FRII sources with green crosses.}
    \label{lobe-pair_metrics_fig}
\end{figure*}

As before with the single-lobe calculations, Figure~\ref{lobe-pair_metrics_fig} shows the pairwise combinations of the source metrics (ABL, $A_{S,p}$, $A'_{S,p}$, $A_{R,p}$, and $A'_{R,p}$), again with the normalised KDE along the diagonal. We can again see that the EMU and 3CRR sources occupy the same areas of parameter space, with differentiation between FRI and FRII sources in each of the asymmetry metrics. FRIIs tend to be more asymmetric than the FRIs. This is likely due to the different structure around the hotspots in each of the radio lobes produced from the higher power jets than in FRI sources. Further, the brightest parts of a FRI source tend to be near the core, where the jets are still highly collimated. This, in conjunction with the limited number of resolution elements for some sources (see Section~\ref{Asymm_analysis}), results in lower asymmetry for the FRI sources. The features observed when comparing the shape and ratio asymmetries for the source asymmetry calculations are notably similar to those seen in the single-lobe asymmetry results. We expect both lobes of a radio source to have similar shape and structure when produced from jets of the same power and in the same environment. For the small number of cases where the environment is not consistent for both lobes (e.g., if the galaxy is on the edge of a cluster), than we expect to see higher asymmetry between lobes. We do however see some variation when comparing the flux-weighted ($A_S$ and $A_R$) and normalised versions ($A'_S$ and $A'_R$) of the asymmetry metrics. Looking at the KDEs for each, we can see that the flux-weighted calculations show a wider spread of values than the normalised ones. This is likely a direct result of the difference in brightness between the two lobes. This difference could be from jets with differing power or a projection effect with the emission of the closer lobe being brighter. This analysis supports the findings of \citet{rodman2019}, who studied the relationship between length asymmetries of radio lobe and the environment properties of the host AGN. They found that the FRI type sources tended to have smaller asymmetries than FRII sources, and that there was a stronger anti-correlation between length ratio and ambient galaxy density than with lobe luminosity and environment. However, we acknowledge that this study had a limited sample size (6 FRI, 16 FRII, 1 hybrid), with limited positive results for FRIs. Our findings are also consistent with \citet{yatesjones2021}.

\subsubsection{Angle Between Lobes}

\begin{figure}[h]
    \centering
    \includegraphics[width=\linewidth]{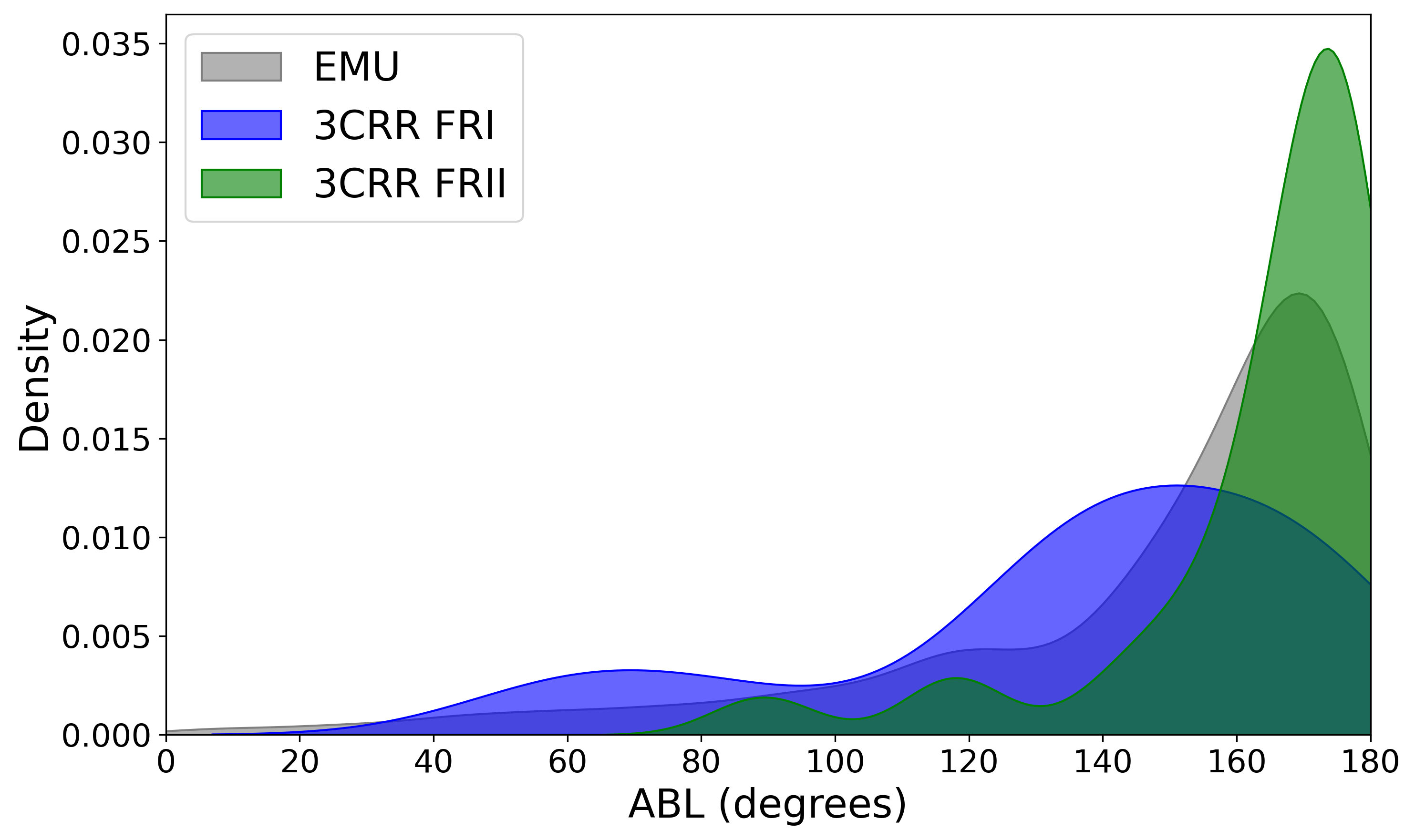}
    \caption{Enlarged version of the ABL KDE sub-panel from Figure~\ref{lobe-lobe_metrics_fig}. EMU sources are represented in grey, and 3CRR sources colour-coded by their FR classification: FRI sources are shown in blue and FRII sources in green.} 
    \label{ABL_KDE}
\end{figure}

Figure~\ref{ABL_KDE} shows that the majority of EMU and FRII sources show little to no bending. There is a distinct separation between FR types at an ABL of roughly $160^\circ$, with FRIs tending to be more bent than FRIIs. Specifically, we find that $53.21\%$ of EMU sources and $76.36\%$ of 3CRR FRII sources have an ABL~$>150^\circ$, with only $29.41\%$ of FRI sources with an ABL~$>150^\circ$. This is to be expected as FRIs have lower power and are often found in higher density environments (like galaxy clusters) than FRIIs. In addition to the lower jet power, FRIs also tend to have slower jet speed. The jets are then more likely to be slowed down through entrainment \citep[e.g.,][]{laingbridle2014}, therefore making them easier to bend \citep{banfield_radio_2015} than FRII sources.

\subsection{Data Clustering in Parameter Space}

To confirm that the metrics presented in this work are connected with physically meaningful aspects of radio galaxy morphology, we used Gaussian Mixture Modelling \citep[GMM;][]{vanderplas_data}, via the \texttt{sklearn.mixture} \texttt{GaussianMixture} package. GMM models the data as a probabilistic mixture of Gaussian distributions, where each cluster is characterised by a mean and covariance structure rather than a strict threshold in any single metric \citep[see e.g., ][]{bishop_pattern}. This allows the clustering process to account for both individual contributions and correlations between metrics, meaning that the separation of clusters emerges from their joint distribution rather than from any one dominant parameter. In this way, GMM can identify if there are multiple populations within a dataset. Before applying GMM, we normalised all metrics to a standard scale spanning a range of $-1$ to $1$ with a mean of $0$, to ensure that each feature contributes equally to the clustering process.

In order to compare against FRI and FRII populations, we specified the GMM to produce two clusters, treating the $ABCDE$ and source metrics as features. We did GMM clustering with the single-lobe metrics ($ABCDE$) and source metrics separately. This was because the source measurements contain half as many data points as the single-lobe measurements (i.e., one measurement per source rather than one per lobe). Doubling the source metrics to match the number of single-lobe measurements would artificially inflate their influence in the clustering process, making subsequent interpretation more ambiguous.

\begin{figure}[h!]
    \centering
    \includegraphics[width=\linewidth]{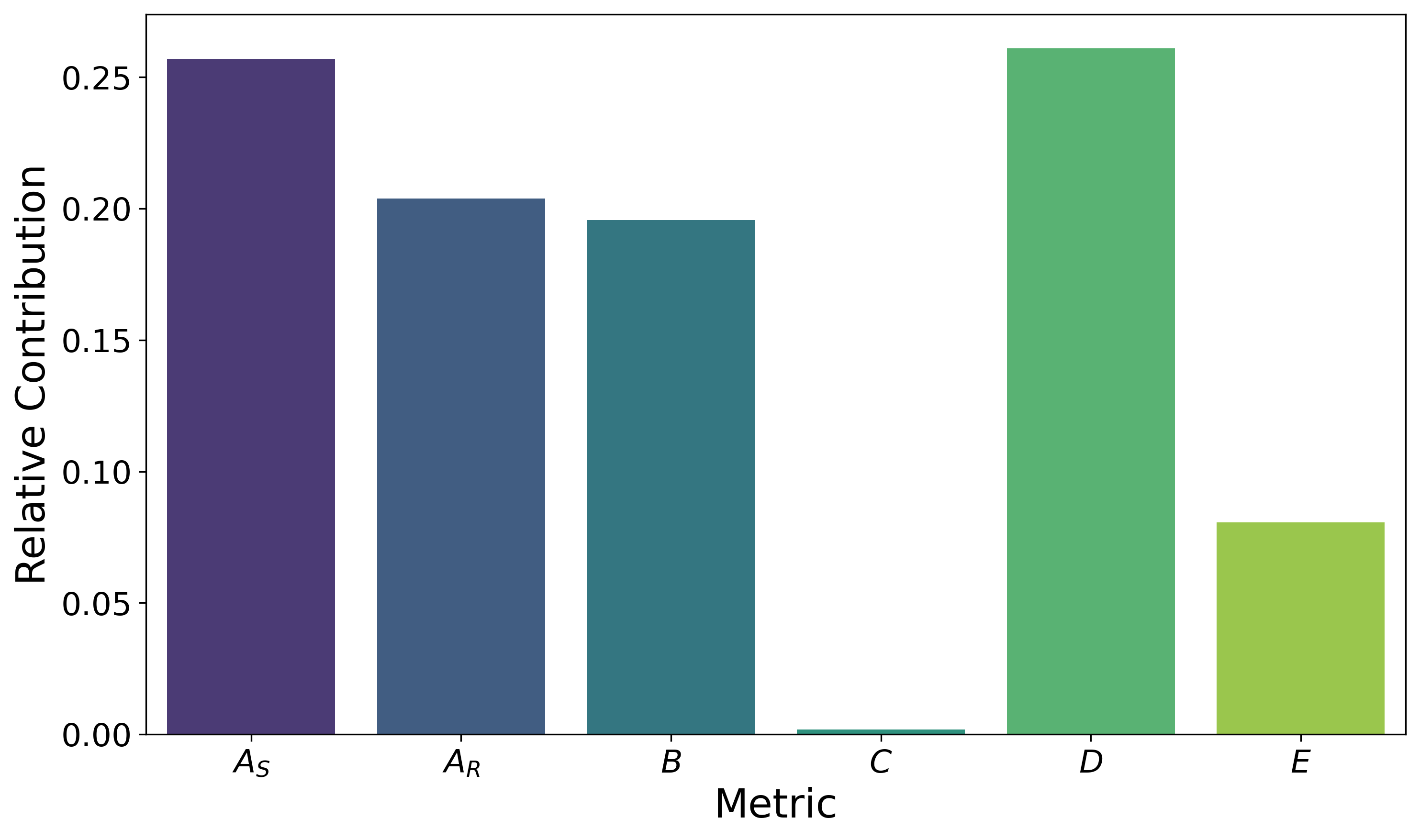}
    \caption{Comparison of the relative contributions of single-lobe metrics in distinguishing between GMM clusters.}
    \label{single_weights_fig}
\end{figure}

\begin{figure}[h!]
    \centering
    \includegraphics[width=\linewidth]{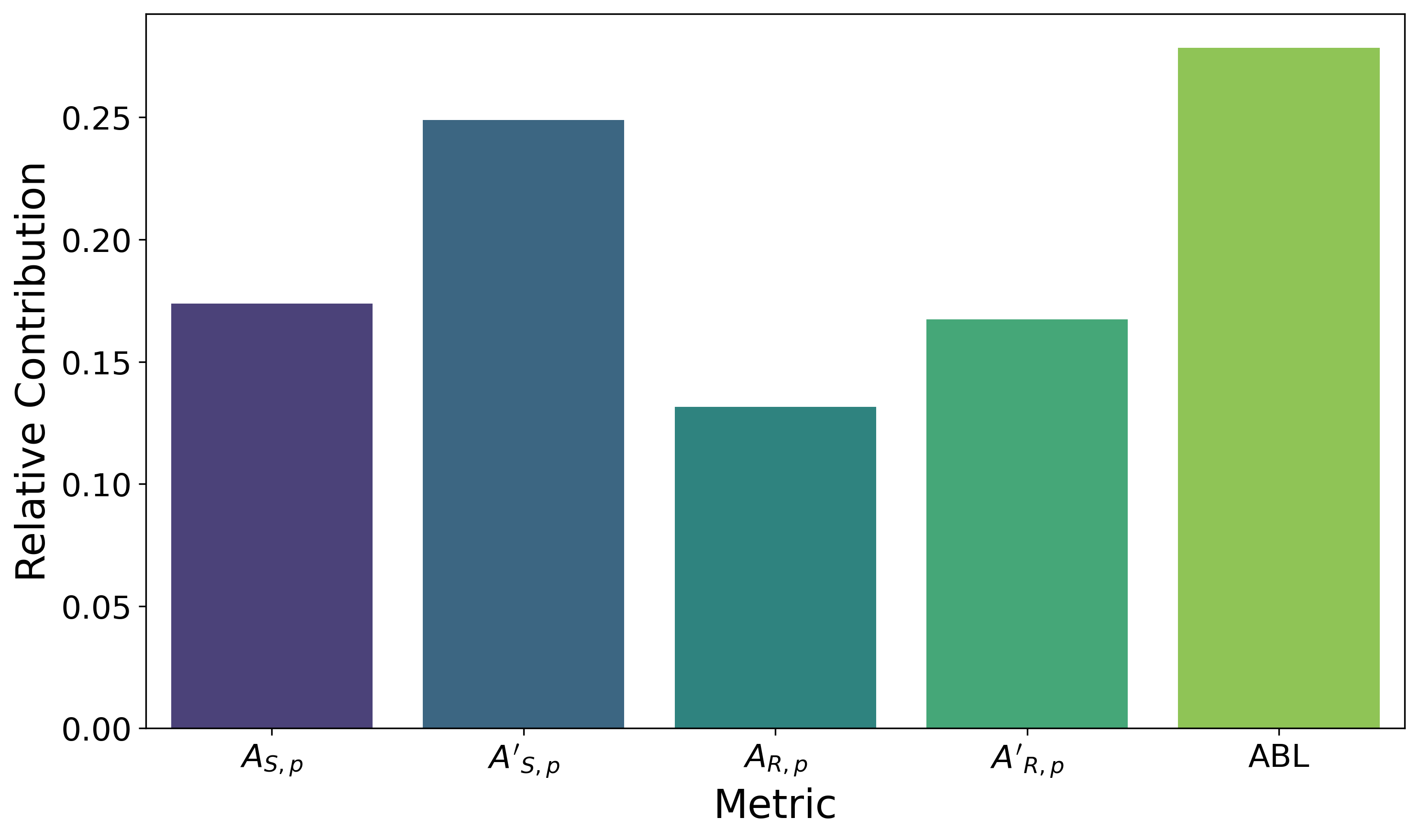}
    \caption{Comparison of the relative contributions of source metrics in distinguishing between GMM clusters.}
    \label{paired_weights_fig}
\end{figure}

Apart from a mean vector and a covariance matrix, GMM also returns a mixture coefficient (weight) that reflects the prior probability of a data point belonging to that component. These mixture weights are not associated with individual features but instead represent the relative prevalence of each Gaussian component in the data. To explore how each feature (or metric in this case) contributes to the separation between clusters, we examined the differences between means of each feature across the two clusters. This provides a simple estimate of which features vary most between clusters, offering insight into their relative contribution to the clustering structure. Although this is not a formal measure of feature importance, it allows us to visualise which metrics show the strongest differences between clusters. This is explored for the single-lobe metrics in Figure~\ref{single_weights_fig}. We can see that the metrics $A_S$, $A_R$, $B$, and $D$ have the highest influence in creating the clusters, with $C$ and $E$ showing little to no contribution. For the source metrics (Figure~\ref{paired_weights_fig}), metrics $A_{S,p}$, $A'_{S,p}$ and ABL, have the highest influence in creating the clusters.

\begin{figure*}[h!]
    \centering
    \includegraphics[width=\linewidth]{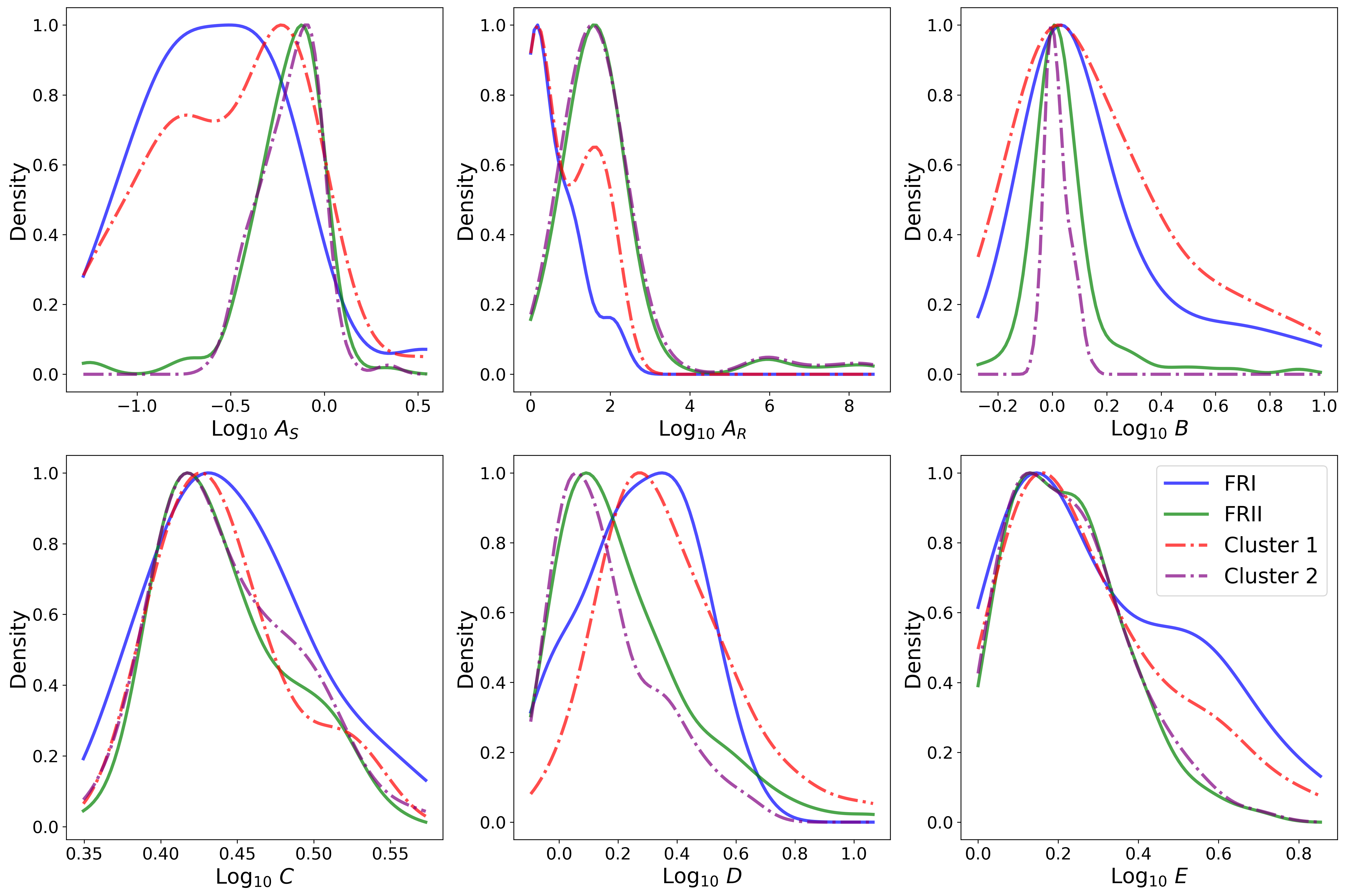}
    \caption{Comparison of the distributions of the GMM clusters with the FR distributions of the single-lobe metrics from Figure~\ref{lobe-lobe_metrics_fig}. Blue and green lines correspond to the distributions of FRI and II sources respectively. Red and purple dashed lines correspond to GMM clusters 1 and 2 respectively. Note how for $A_R$ and $D$, the distributions are quite different for FRIs and FRIIs, and are well-matched to the Cluster 1 and 2 distributions, respectively.}
    \label{single_cluster_FR_fig}
\end{figure*}

\begin{figure*}[h!]
    \centering
    \includegraphics[width=\linewidth]{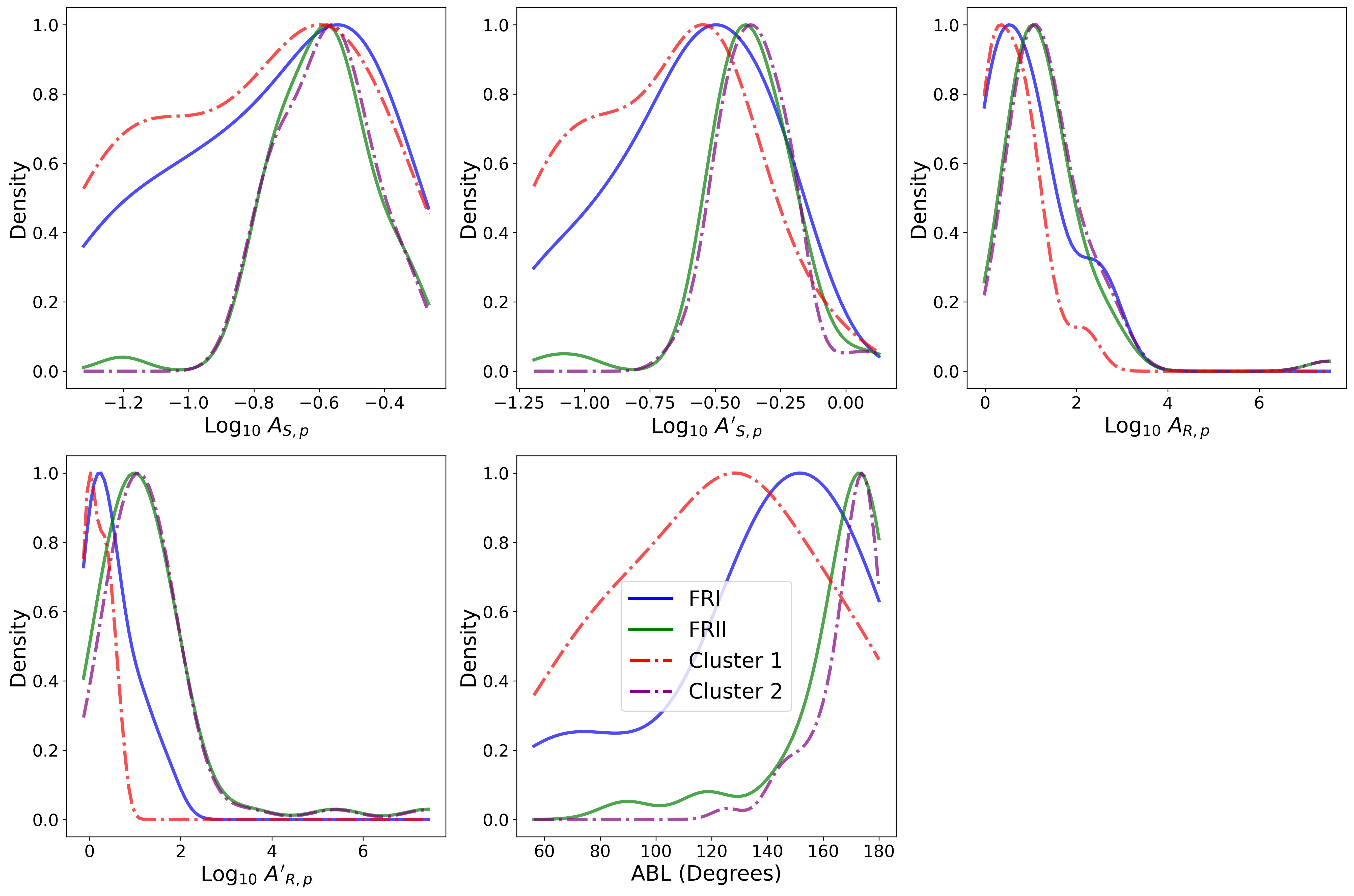}
    \caption{Comparison of the distributions of the GMM clusters with the FR distributions of the source metrics from Figure~\ref{lobe-lobe_metrics_fig}. Blue and green lines correspond to the distributions of FRI and II sources respectively. Red and purple dashed lines correspond to GMM clusters 1 and 2 respectively. Again, note how for $A'_{S,p}$, $A'_{R,p}$ and ABL, the difference in distributions between FRIs and FRIIs, that are well approximated by the distributions of Clusters 1 and 2, respectively.}
    \label{paired_cluster_FR_fig}
\end{figure*}

Figure~\ref{single_cluster_FR_fig} compares the distributions of the GMM clusters with the FR distributions of the single-lobe metrics from Figure~\ref{lobe-lobe_metrics_fig}. The GMM clusters match closely with the FR distributions (cluster 1 with FRI, cluster 2 with FRII) in our metric space. The cluster and FR distributions are particularly close in the $A_R$, and $D$ metrics, reflecting the relative importance of these metrics in the clustering process. The clusters made with the source metrics also align closely with FR distributions (Figure~\ref{paired_cluster_FR_fig}), particularly in $A_{R,p}$ and $A'_{R,p}$. This suggests that while the GMM relies more heavily on $A_{S,p}$ and $A'_{S,p}$ to separate the data probabilistically, the actual FR classification differences manifest more strongly in the $A_{R,p}$ and $A'_{R,p}$ distributions. The results highlighted in Figures~\ref{single_cluster_FR_fig} and \ref{paired_cluster_FR_fig} confirm that both the $ABCDE$ and source metrics are physically motivated and are not arbitrary shape analysis calculations.

To assess the performance of the clustering in distinguishing between FR morphologies, we compute completeness and reliability for each class. We define completeness as the fraction of sources of a given FR type that are correctly assigned to the corresponding cluster. Specifically for FRI sources, completeness is defined as the number of FRI sources in the FRI-predominant cluster divided by the total number of FRI sources. Reliability quantifies how pure a given cluster is with respect to an FR type. We define it as the fraction of sources in the FRI-predominant cluster that are actually classified as FRI. The same definitions apply to FRII sources. A high completeness value indicates that most FRI (or FRII) sources are correctly assigned, while a high reliability value suggests that the cluster is largely free from contamination by the other FR type. These values, for both single-lobe and source metrics, are compiled in Table \ref{tab:comp_reliability}. On average, the clustering is more successful at grouping the FRII sources together, likely due to larger number of FRIIs in our sample.

\begin{table}[h!]
    \centering
    \begin{tabularx}{\linewidth}{l *{4}{>{\centering\arraybackslash}X}}
        \hline
        \hline
        \textbf{Metric Type} & \multicolumn{2}{c}{\textbf{Completeness (\%)}} & \multicolumn{2}{c}{\textbf{Reliability (\%)}} \\
        & Cluster 1 & Cluster 2 & Cluster 1 & Cluster 2 \\
        \hline
        Single-Lobe  & 70.6 & 78.2 & 50.0 & 89.6 \\
        Source       & 58.8 & 89.1 & 62.5 & 87.5 \\
        \hline
        \hline
    \end{tabularx}
    \caption{Completeness and reliability of GMM clusters mapped to FR classifications, using the single-lobe ($ABCDE$) and source metrics (ABL, $A_{S,p}$, $A'_{S,p}$, $A_{R,p}$, $A'_{R,p}$).}
    \label{tab:comp_reliability}
\end{table}

To further quantify the similarity between distributions of the FR types with their respective clusters, we perform the Kolmogorov–Smirnov \citep[KS,][]{kolmogorv_1933, smirnov_1948} test. Specifically we use the two-sample KS test, which takes the null hypothesis that two distributions come from the same population. If the $p$-value of the KS test is small ($p<0.05$), we can conclude that the two distributions were sampled from different underlying populations. We used the KS test to compare FRI with cluster 1 and FRII with cluster 2, compiling the resulting $p$-values in Table~\ref{KS_test_results}. We can see that for all metrics $p>>0.05$, consistent with the null hypothesis, that both are drawn from the same population. This further reinforces that the sources in cluster 1 are consistent with the properties of FRIs, and those in cluster 2 with FRIIs.

\begin{table}[h]
\centering
\caption{Kolmogorov--Smirnov test results comparing FRI sources with Cluster 1 and FRII sources with Cluster 2 for both the single-lobe metrics ($ABCDE$), and source metrics.}
\label{KS_test_results}
\begin{tabularx}{\linewidth}{lXX}
\hline
\hline
\textbf{Metric} & \textbf{FRI vs Cluster 1} & \textbf{FRII vs Cluster 2} \\
\hline
$A_S$ & $p = 3.840\times10^{-1}$ & $p = 1.000$ \\
$A_R$ & $p = 1.160\times10^{-1}$ & $p = 1.000$ \\
$B$ & $p = 8.280\times10^{-1}$ & $p = 7.020\times10^{-1}$ \\
$C$ & $p = 9.130\times10^{-1}$ & $p = 1.000$ \\
$D$ & $p = 5.410\times10^{-1}$ & $p = 4.240\times10^{-1}$ \\
$E$ & $p = 9.190\times10^{-1}$ & $p = 1.000$ \\[1ex]
\hline
\rule{0pt}{3mm}$A_{S,p}$ & $p = 9.420\times10^{-1}$ & $p = 1.000$ \\
$A'_{S,p}$ & $p = 9.070\times10^{-1}$ & $p = 1.000$ \\
$A_{R,p}$ & $p = 6.640\times10^{-1}$ & $p = 9.990\times10^{-1}$ \\
$A'_{R,p}$ & $p = 3.650\times10^{-1}$ & $p = 9.920\times10^{-1}$ \\
ABL & $p = 2.790\times10^{-1}$ & $p = 9.520\times10^{-1}$ \\
\hline
\hline
\end{tabularx}
\end{table}

While we do not expect a perfect 1:1 correlation between the GMM clusters and the FR classifications, the observed correspondence between them in this simple comparison strongly reinforces the value of quantifying radio galaxy structure through the $ABCDE$ and source metrics. The fact that these metrics can probabilistically separate sources into distinct clusters that correspond to known FR types (to some degree) demonstrates the utility of these metrics as a flexible tool for classifying and quantifying radio galaxy morphology.

\begin{table}[h]
\centering
\caption{Kolmogorov--Smirnov test results comparing FRI sources with Cluster 1 and FRII sources with Cluster 2 for both the single-lobe metrics ($ABCDE$), and source metrics.}
\label{KS_test_results}
\begin{tabularx}{\linewidth}{lXX}
\hline
\hline
\textbf{Metric} & \textbf{FRI vs Cluster 1} & \textbf{FRII vs Cluster 2} \\
\hline
$A_S$ & $p = 3.840\times10^{-1}$ & $p = 1.000$ \\
$A_R$ & $p = 1.160\times10^{-1}$ & $p = 1.000$ \\
$B$ & $p = 8.280\times10^{-1}$ & $p = 7.020\times10^{-1}$ \\
$C$ & $p = 9.130\times10^{-1}$ & $p = 1.000$ \\
$D$ & $p = 5.410\times10^{-1}$ & $p = 4.240\times10^{-1}$ \\
$E$ & $p = 9.190\times10^{-1}$ & $p = 1.000$ \\
\hline
$A_{S,p}$ & $p = 9.420\times10^{-1}$ & $p = 1.000$ \\
$A'_{S,p}$ & $p = 9.070\times10^{-1}$ & $p = 1.000$ \\
$A_{R,p}$ & $p = 6.640\times10^{-1}$ & $p = 9.990\times10^{-1}$ \\
$A'_{R,p}$ & $p = 3.650\times10^{-1}$ & $p = 9.920\times10^{-1}$ \\
ABL & $p = 2.790\times10^{-1}$ & $p = 9.520\times10^{-1}$ \\
\hline
\hline
\end{tabularx}
\end{table}

\subsection{Computational Efficiency}

To estimate how computationally efficient our algorithm is, we timed how long it took to compute both the $ABCDE$ and source metrics for all images in the EMU and 3CRR datasets. These computations were performed on an Apple MacBook Pro with an M$4$ chip and $16$ GB of RAM, running in a Python notebook. We repeated this process $10$ times for both datasets, then took the average. We obtained an average computation time of $10.56$ seconds for the $480$ EMU images, averaging $0.022$ seconds per image. For the $72$ 3CRR images, we found an average computation time of $5.02$ seconds, averaging $0.070$ seconds per image. The average number of pixels in the EMU images was $3,596.88$, whereas for the higher-resolution 3CRR images, the average number of pixels was $115,729.85$. The difference in computation time between EMU and 3CRR is therefore likely dependent on the number of pixels in the image. This demonstrates that the algorithm performs better than linear, as a roughly $30\times$ increase in the number of pixels resulted in only a $\sim\! 3\times$ increase in computation time on average. This efficiency highlights the algorithm’s scalability, making it feasible for large-scale surveys like EMU. Further optimisations, such as parallelising computations or leveraging GPU acceleration, may improve performance even further.

\section{Conclusions}

\begin{figure}[h]
    \centering
    \includegraphics[width=\linewidth]{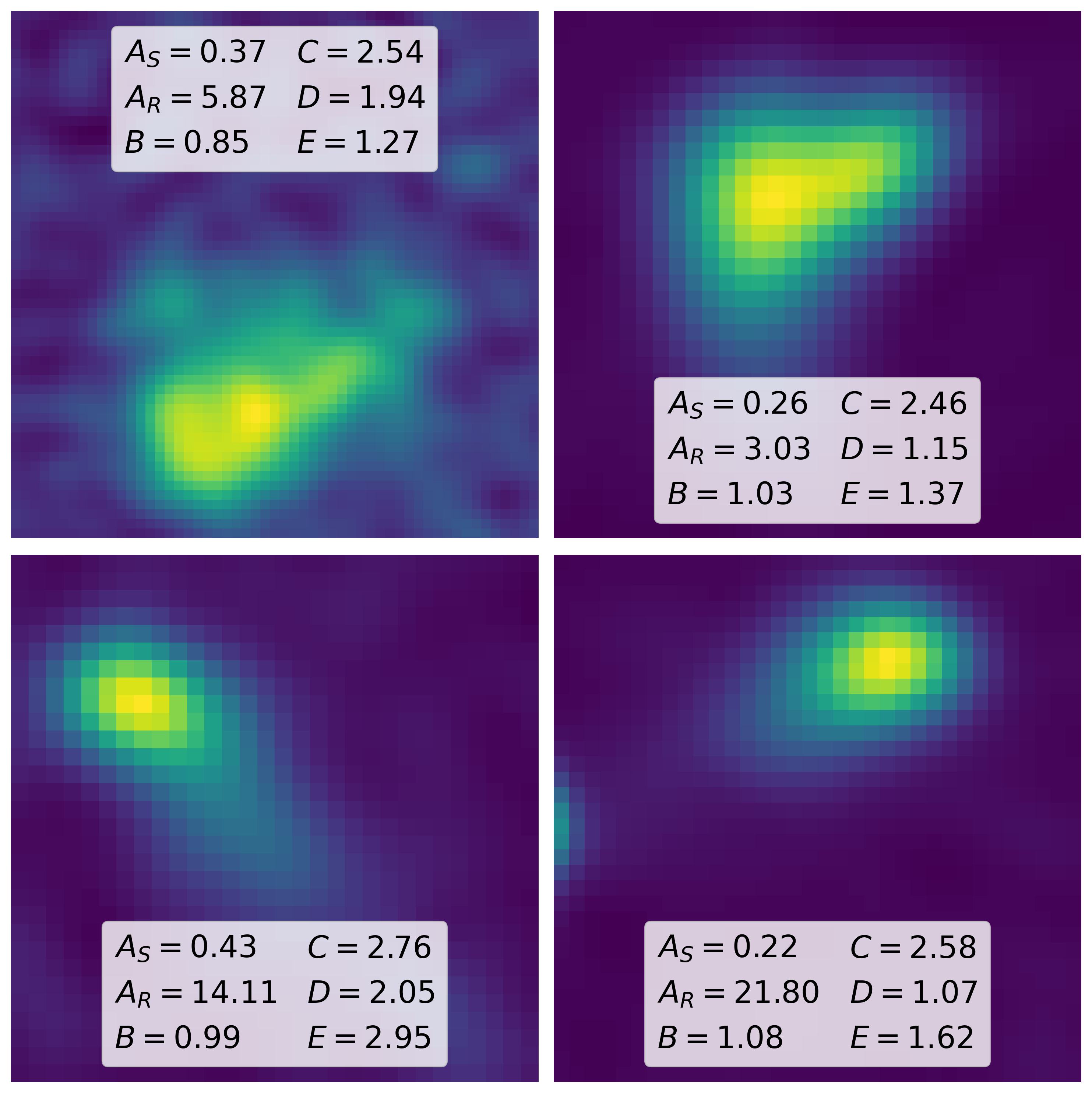}
    \caption{Multi-panel comparison of single lobes from different sources in the EMU-PS. Visual differences in each lobe structure are highlighted by different values of the $ABCDE$ metrics in the legend.}
    \label{abcde_metrics}
\end{figure}

We have introduced a series of quantitative metrics to describe the structure of a single radio galaxy lobe, and of the whole DRAGN. These include two asymmetry calculations, blurriness, concentration, disorder, and elongation ($ABCDE$), asymmetry and angle between lobes (source metrics). We developed this series of metrics metrics to be applied to large radio datasets, such as those that will be produced by EMU, and surveys from MeerKAT, LOFAR, SKA (in future), in an automatic and more efficient way than current machine learning approaches. We applied our metrics to two datasets: $480$ sources from the EMU-PS and $72$ sources from the 3CRR catalogue. We propose that these metrics offer the prospect of deeper insights into both the structure and astrophysical processes shaping the radio galaxy lobes than the earlier Fanaroff-Riley morphology classification system (illustrative examples are shown in Figure~\ref{abcde_metrics}). These metrics can also be incorporated into the \textit{\#tag} system proposed by \citet{rudnick_radio_2021}. While the current set of metrics presented in this work provides an intuitive and effective starting point for quantifying complex radio structures, it is by no means exhaustive. 

We found that the our metrics are largely independent of each other with Spearman rank correlation coefficient of $-0.01 < \rho < 0.3$ for all metric combinations except $A_S$-$A_P$ and $D$-$E$, which had a $\rho$ of $0.72$ and $0.57$, respectively. While the quantitative values for some of these metrics will change with the resolution (see Section ~\ref{res_dependence}), their qualitative performance remains consistent. The minimum image area of $20$ beams for the EMU sources resulted in them being sufficiently resolved for the metric results to be comparable with the 3CRR data, evidenced by the metric results of both datasets occupying the same areas of parameter space. Further, a source with higher surface brightness will be less susceptible to noise biases and therefore should produce more robust results. We therefore acknowledge that this series of metrics will tend to have better performance on well resolved sources with high signal-to-noise, on average. In both single-lobe and source calculations, we find that the FRII types tend to have higher asymmetry than the FRIs, likely due to the differences in radio power and the structural features around hotspots. This could also be due to a potential selection effect. For FRI sources, a lack of sensitivity will mean that only the jets will be seen, which is expected to be symmetric before the jets becomes disrupted, whereas the edge-brightened nature of FRII sources allow us to see more complex structure of the lobe. Additionally, the flux-weighted asymmetry metrics in the source calculations show more scatter when compared to the normalised versions, indicating some brightness differences between the lobes. 

We performed clustering analysis on the 3CRR data using Gaussian mixture modelling (GMM) with the single-lobe $ABCDE$ metrics and source metrics separately. We find the distributions of the GMM clusters align with the distributions of the FR classifications in our parameter space, reinforcing the physical relevance of these metrics. While we do not expect a perfect one-to-one correspondence between the clusters and FR types, the observed overlap highlights the effectiveness of this quantitative approach. Notably, the clustering was more successful in grouping FRII sources, likely due to the larger sample size in our data. 

We find that ABL is a relatively important source metric for radio galaxy morphology. The ABL measurements show a clear distinction between FR types, with FRI sources being significantly more bent than FRII sources. This is consistent with the expectation that FRI sources, which are typically found in higher-density environments and often have lower jet power, experience greater interactions with the surrounding medium. While ABL does not directly fit into the single-lobe $ABCDE$ framework, it plays a significant role in characterising source morphology. Additionally, our clustering analysis shows that ABL, along with $A_{S,p}$ and $A'_{S,p}$, has a strong influence on how clusters are assigned through GMM. ABL is therefore a key parameter in defining the source structure, highlighting the need to consider both single-lobe and source metrics when analysing radio galaxy morphology.

This series of metrics can be particularly valuable in the era of large-scale surveys, where millions of sources will need to be analysed in an automated and efficient manner. By capturing key aspects of radio lobe structure, our metrics can reveal critical information about AGN and the astrophysical processes driving jet formation, particle acceleration, and interactions between the radio source and the surrounding environment. Further, the metrics presented in this work are agnostic to the changing emphases in radio galaxy interpretation. They do not depend on specific morphological labels, such as FRI or FRII. As a result, they can be used to efficiently characterise large surveys without being constrained by predefined classifications. Once characterised, different combinations of metrics can then be applied to identify new and interesting types of sources.

\section{Future work}

Future work will focus on refining the implementation of the metrics, to reduce the effect of sources with limited sampling. This could be done by implementing a region-based criteria for lobe detection, instead of using the brightest pixel, by detecting extended regions of high flux and using contour-based methods to locate the true lobes. This could enable, for example, the modification of the definitions of asymmetry metrics to span the full extent of a lobe plus jet, providing a more astrophysically intuitive measurement. This work does not explicitly exclude low surface brightness sources in order to investigate how the metrics perform on a range of source types. Further investigation, perhaps through binning sources by brightness, can provide a more robust understanding of how the metrics perform with varying source brightness. The inclusion of new metrics into our current framework may also improve the ability to quantify complex radio structures, offering the prospect of deeper insights about radio galaxy morphology. Applying these metrics to a dataset with a larger sample size, such as MiraBest\footnote{\hyperlink{https://doi.org/10.5281/zenodo.4288837}{https://doi.org/10.5281/zenodo.4288837}} \citep{porter_mirabest_2023} is also a crucial step in the testing of the metrics.  

A key direction will be further exploring the relationship between the the metrics presented in this work and the underlying astrophysical processes. Linking these metrics to characteristics such as AGN properties, galaxy number density, and host galaxy properties (such as mass, colour, optical morphology, among others), will enable a more comprehensive understanding of the forces shaping radio galaxies. 

Multi-frequency analysis of radio sources will be crucial in future studies. By examining how the metric values vary across different radio frequencies, we can gain a clearer picture of how various physical processes, such as different emission mechanisms, environmental impacts, and jet composition and its dynamics, manifest at different frequencies. This will not only enhance our understanding of radio galaxies and their classifications but also provide valuable insights into the broader mechanisms driving galaxy evolution.

\begin{acknowledgement}
This scientific work uses data obtained from Inyarrimanha Ilgari Bundara, the CSIRO Murchison Radio-astronomy Observatory. We acknowledge the Wajarri Yamaji People as the Traditional Owners and native title holders of the Observatory site. CSIRO’s ASKAP radio telescope is part of the Australia Telescope National Facility (https://ror.org/05qajvd42). Operation of ASKAP is funded by the Australian Government with support from the National Collaborative Research Infrastructure Strategy. ASKAP uses the resources of the Pawsey Supercomputing Research Centre. Establishment of ASKAP, Inyarrimanha Ilgari Bundara, the CSIRO Murchison Radio-astronomy Observatory and the Pawsey Supercomputing Research Centre are initiatives of the Australian Government, with support from the Government of Western Australia and the Science and Industry Endowment Fund.

This paper includes archived data obtained through the CSIRO ASKAP Science Data Archive, CASDA (\url{http://data.csiro.au}).
\end{acknowledgement}

\printendnotes

\bibliography{main}

\appendix

\section{Troublesome Images}\label{trouble_imgs}

\begin{figure}[h]
    \centering
    \includegraphics[width=\linewidth]{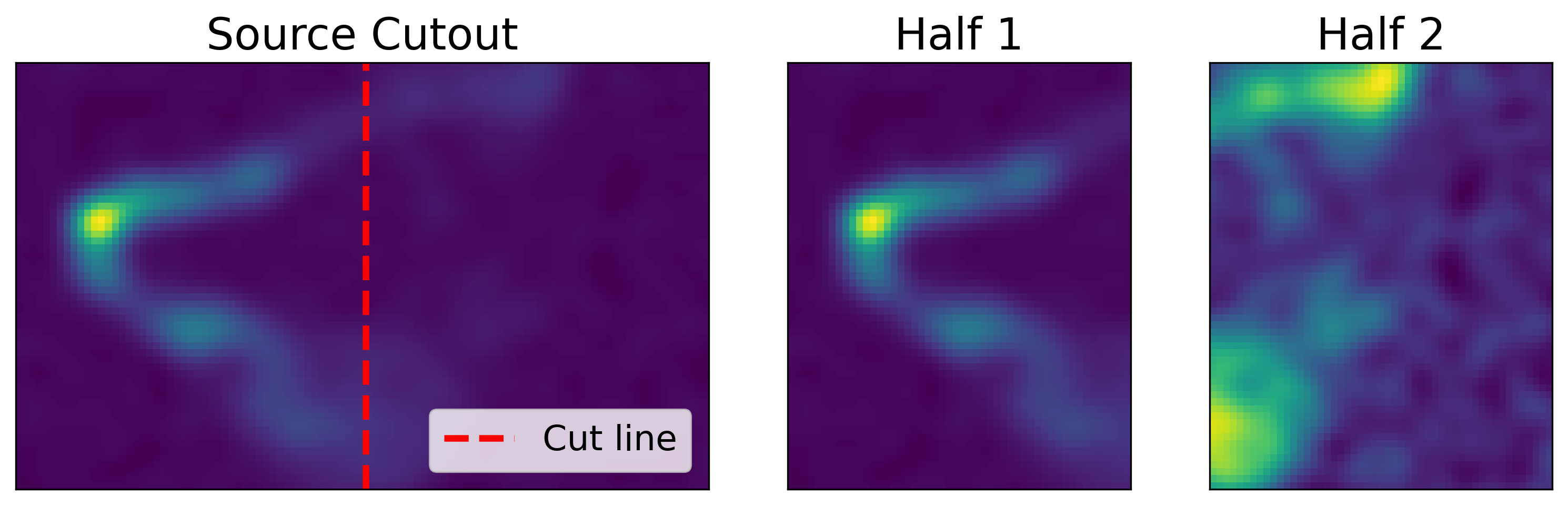}
    \caption{Example of an extremely bent source where both lobes are in one side of the cutout. Red dashed line indicates the midpoint of the major axis of the source image.}
    \label{bent_bad_cut}
\end{figure}

The metric algorithm described above works on a series of assumptions. First, we require that an image is centred in the cutout, so that there is a lobe in each half (cutting the image perpendicular to and half-way along the major axis). This is not true for all sources, for example, those that are extremely bent. For these objects both lobes are in the same half of the cutout (e.g., Figure~\ref{bent_bad_cut}). For such sources, rather than our metrics quantifying some properties of a lobe, we would instead have some measurement of the core and perhaps jets, and either missing both lobes or possibly incorporating one of them. While the metric calculation would return some measurement, it would not be the measurement intended. Accordingly, through visual inspection, extremely bent sources were omitted from our analysis.

\begin{figure}[h]
    \centering
    \includegraphics[width=\linewidth]{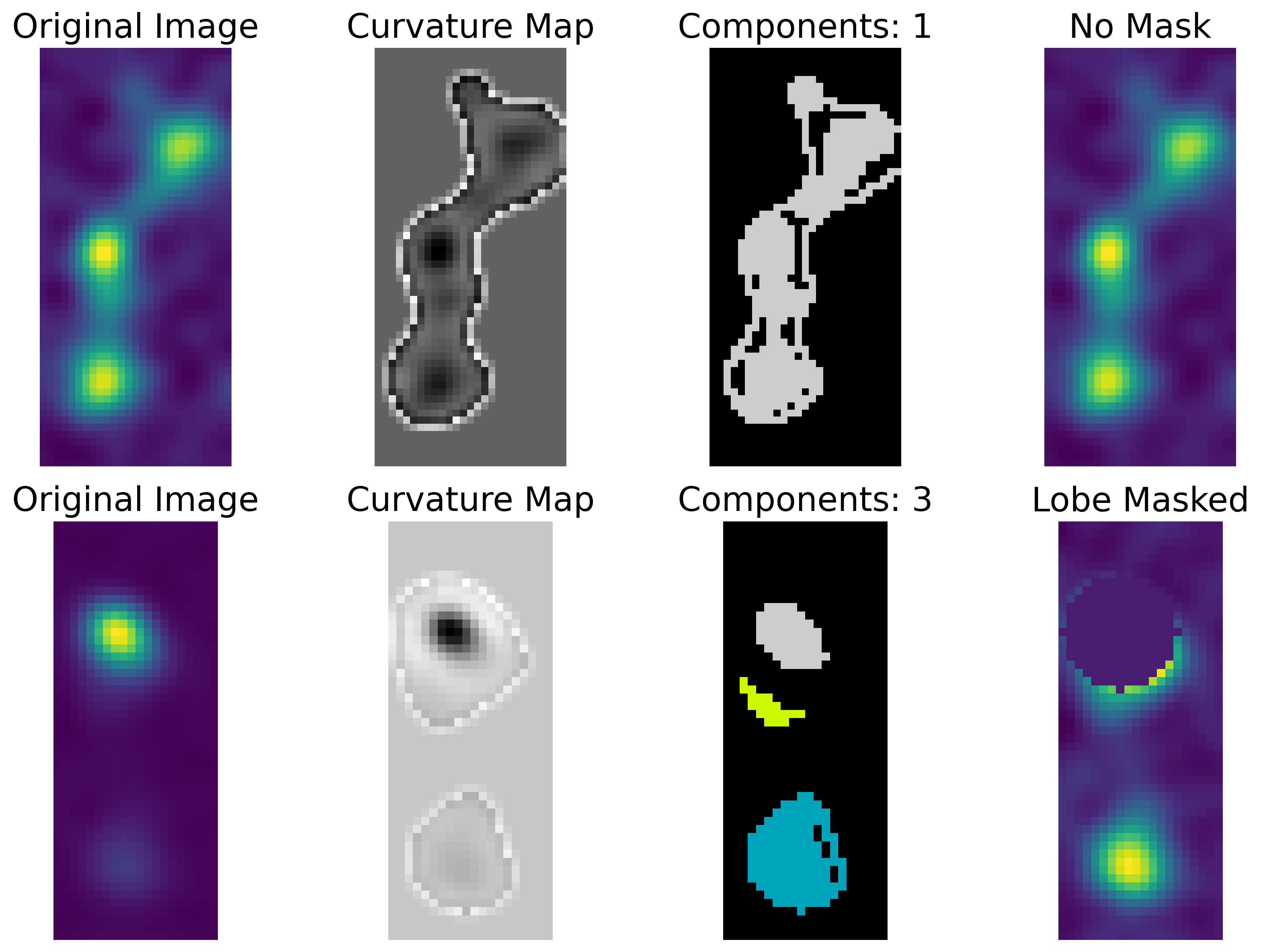}
    \caption{Illustrative examples of when the core-removal process fails. Top row: number of source components was underestimated, and no mask was applied. Bottom row: number of source components was overestimated, and a lobe was masked.}
    \label{num_comps_wrong}
\end{figure}

Second, we ideally want sources that are not core-dominated, so that the core is less luminous than the lobes. This is because our algorithm uses the brightest pixel as the base for many of the metric operations ($A_S$, $A_R$, $B$, $C$, for example). As mentioned in Section~\ref{preprocessing}, the core removal process we implemented did not always work perfectly. For example, there were cases where an erroneous number of components were detected. If the number of components was under-estimated the core would not be masked out, and if it was over-estimated a lobe could be masked out (see Figure~\ref{num_comps_wrong} for examples of each case). Both scenarios again would lead to the metric measurements not reflecting the actual lobe structure. Such sources were excluded from our analysis.

\section{Metric differences between lobes}

\begin{figure}
    \centering
    \includegraphics[width=\linewidth]{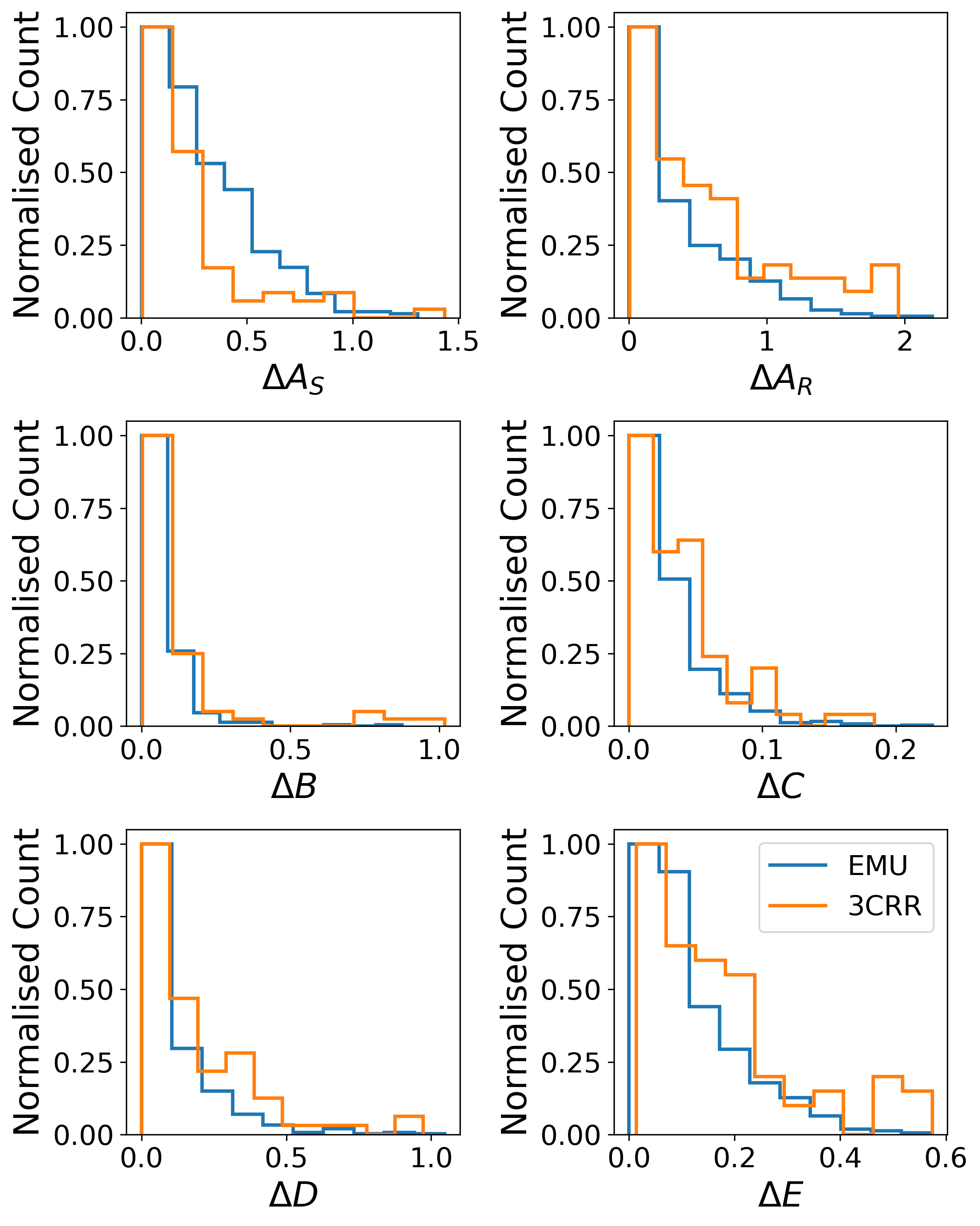}
    \caption{Normalised histograms of absolute differences ($\Delta x$) between single lobe metrics for EMU (blue) and 3CRR (orange) sources.}
    \label{diff_between_lobes}
\end{figure}

To quantify how the metrics differ between lobes, we simply take the absolute difference of the log$_{10}$ values the metric values for each lobe:

\begin{equation}
    \Delta x = |log_{10}(x_1) - log_{10}(x_2)|~,
\end{equation}

\noindent where $x_1$ and $x_2$ are the metric values of lobe $1$ and $2$ respectively. In Figure~\ref{diff_between_lobes} we show the range of differences between each lobe, across each of the $ABCDE$ metrics. We calculate the percentage of sources with $\Delta x < 0.3$ (Table \ref{delta_x_percentages}). We can see for both EMU and 3CRR sources, the majority of lobe measurements are within a factor of two of each other for each of the $ABCDE$ metrics. The $A_R$ metric appears to show the greatest degree of difference suggesting that it could be the most sensitive to variations between lobes.

\begin{table}[h]
\centering
\caption{Percentage of sources with $\Delta x<0.3$ for both EMU and 3CRR datasets.}
\label{delta_x_percentages}
\begin{tabularx}{\linewidth}{>{\centering\arraybackslash}X>{\centering\arraybackslash}X>{\centering\arraybackslash}X}
\hline
\hline
\boldmath{$\Delta x$} & \textbf{EMU (\%)} & \textbf{3CRR (\%)} \\
\hline
$\Delta A_S$ & 59.9 & 76.4 \\
$\Delta A_R$ & 58.2 & 40.3 \\
$\Delta B$ & 97.6 & 91.2 \\
$\Delta C$ & 100.0 & 100.0 \\
$\Delta D$ & 90.2 & 75.0 \\
$\Delta E$ & 93.5 & 83.3 \\
\hline
\hline
\end{tabularx}
\end{table}

\end{document}